\documentclass[aip,reprint]{revtex4-1}
\pdfoutput=1
\usepackage{amssymb}
\usepackage{graphicx}
\usepackage{subfloat}
\usepackage{epsfig}
\usepackage{multirow}
\usepackage{epstopdf}
\usepackage{gensymb}
\usepackage{nicefrac}
\usepackage{mathtools}
\usepackage{color}
\usepackage{subfiles}
\usepackage{blindtext}
\usepackage{soul,xcolor}
\usepackage{fontenc}
\usepackage{romannum}

\graphicspath{ {./} }

\usepackage[left]{lineno}

\usepackage[utf8]{inputenc}
\usepackage{geometry}
\usepackage{cleveref}
\usepackage{amsmath}
\usepackage{caption}
\usepackage{xr}
\usepackage{titlesec}
\titlespacing*{\section} {0pt}{3.5ex plus 1ex minus .2ex}{2.3ex plus .2ex}
\titlespacing*{\subsection} {0pt}{3.25ex plus 1ex minus .2ex}{1.5ex plus .2ex}

\draft 

\begin{document}

\preprint{}

\title{Range-Separated Hybrid Functionals for Mixed Dimensional Heterojunctions: Application to Phthalocyanines/MoS$_2$} 

\author{Qunfei~Zhou}
\affiliation{Materials Research Science and Engineering Center, Northwestern University, Evanston, IL 60208, USA}
\affiliation{Center for Nanoscale Materials, Argonne National Laboratory, Argonne, IL 60439, USA}

\author{Zhen-Fei Liu}
\email{zfliu@wayne.edu}
\affiliation{Department of Chemistry, Wayne State University, Detroit, MI 48202, USA}

\author{Tobin J. Marks}
\email{t-marks@northwestern.edu}
\affiliation{Department of Chemistry and Department of Materials Science and Engineering, Northwestern University, Evanston, IL 60208, USA}
\affiliation{Materials Research Science and Engineering Center, Northwestern University, Evanston, IL 60208, USA}

\author{Pierre~Darancet}
\email{pdarancet@anl.gov}
\affiliation{Center for Nanoscale Materials, Argonne National Laboratory, Argonne, IL 60439, USA}
\affiliation{Northwestern Argonne Institute for Science and Engineering, Evanston, IL 60208, USA}

\date{\today}

\begin{abstract}
We analyze the electronic structure and level alignment of transition-metal phthalocyanine (MPc) molecules adsorbed on two-dimensional MoS$_2$ employing density functional theory (DFT) calculations. We develop a procedure for multi-objective optimal tuning of parameters of range-separated hybrid functionals in these mixed-dimensional systems. Using this procedure, which leads to the asymptotically-correct exchange-correlation potential between molecule and two-dimensional material, we obtain electronic structures consistent with experimental photoemission results for both energy level alignment and electronic bandgaps, representing a significant advance compared to standard DFT methods. We elucidate the MoS$_2$ valence resonance with the transition-metal phthalocyanine non-frontier 3$d$ orbitals and its dependence on the transition metal atomic number. Based on our calculations, we derive parameter-free, model self-energy corrections that quantitatively accounts for the effects of the heterogeneous dielectric environment on the electronic structure of these mixed-dimensional heterojunctions.
\end{abstract}

\maketitle 

\section{Introduction}
Mixed-dimensional van der Waals heterojunctions (MDHJ)~\cite{Hersam2017MDHJ} comprised of two-dimensional (2D) materials and 0D constructs such as molecules and quantum dots exhibit  electronic and optical properties promising for a wide variety of devices including field-effect transistors, sensors, and light-emitting diodes~\cite{Hersam2016NL,Hersam2015NL,Hersam2018NLDevice,Hersam2014Review}. The interfacial coupling in MDHJs gives rise to emergent properties distinct from the ones of their individual components, including photoresponse ~\cite{Amsterdam2019PcMDHJ} and band gaps ~\cite{Quek2019dielectricHJ}. Many of these favorable optoelectronics properties arise as a direct consequence of the electronic level alignment between the different components of the MDHJ~\cite{Hersam2020InSe}. 

As a result of the extreme heterogeneity in the density of states and dielectric screening of these systems, the electronic properties of MDHJs are impacted by numerous competing energy scales, such as the ones associated with local and  non-local electronic correlations, interface dipoles, orbital hybridization, that can lead to significant renormalization of the intrinsic energy levels of each material at these interfaces~\cite{Hersam2020InSe}. Correspondingly, the variety of these energy scales also complicates the theoretical description of the MDHJs  electronic structure. 
In particular, density functional theory (DFT) methods based on Kohn-Sham equations and local exchange and correlation potentials correctly describe the impact of quantum confinement on bandgaps~\cite{Sapori2016Dielectric} and the effect of interface charge transfer, but suffer from self-interaction errors and lack non-local correlations. 
Two families of approaches have been used to correct these two deficiencies: many-body perturbation theory within the GW approximation~\cite{Louie1986GW,Louie2015GW,Galli2016GW} and generalized Kohn-Sham methods using range-separated hybrid functionals~\cite{Baer2009OTRSH,Baer2010OTRSH,Kronik2011OTRSH,Kronik2014OTRSHJCTC}. Range-separated hybrid functionals allow for accurate self-consistent calculations through the generation of a system-dependent set of physical parameters. Previous works have shown that a physically-constraint,  parameter-free tuning of these functionals can lead to quantitatively accurate predictions of the electronic and optical properties of various materials, including organic~\cite{Kronik2013SRSH} and inorganic~\cite{Sahar2020OTSRSH} solids, as well as 2D materials~\cite{Ramasubramaniam2019SRSH2D}. However, the applicability of these functionals and optimization criteria to multi-component and mixed-dimensional systems such as MDHJs remains an open question.

In this work, we use optimally-tuned, screened range-separated hybrid functionals (SRSH) to calculate the electronic structure of experimentally synthesized~\cite{Amsterdam2019PcMDHJ,Chowdhury2017MPcMDHJ} MDHJs comprised of robust, technologically significant metal-free and transition-metal phthalocyanine dye molecules~\cite{Walter2010PcApp,Chen2018PcProp,Lu2016PcRev,Berndt2012PcSTM} (H$_2$Pc and MPc with M=Co, Zn) on monolayer MoS$_2$. We use a classical electrostatic model~\cite{Neaton2006OrgHJ,Liu2017OrgHJ,Neaton2013HJSigma,Quek2019dielectricHJ,Kronik2015OrgHJ} to integrate the dielectric screening effects into the SRSH functional on top of the compliance to Koopmans' theorem~\cite{Kronik2013SRSH} for individual components. We show that the static non-local correlations lead to a $\simeq 2.0$ eV bandgap renormalization for the adsorbed molecules. Moreover, we elucidate the dependence of the MDHJ electronic structure on the non-frontier, $d$-orbitals of the MPc molecules, demonstrating that their hybridization with the MoS$_2$ valence band strongly depends on the center metal atoms, an effect beyond the reach of perturbative corrections to local density functional calculations. 
We obtain electronic bandgaps  in the MDHJ in quantitative agreement with experimental photoemission spectroscopy results~\cite{Mutz2020PcMoS2}. We find a type II band alignment for all three MDHJs with the phthalocyanine highest occupied molecular orbital lying within the bandgap of MoS$_2$. Our work demonstrates, for the first time, the ability of SRSH functionals to describe the electronic structure of MDHJs and emphasizes the importance of including the dielectric screening in the description of these systems. 

\section{Computational Details}
DFT calculations for H$_2$Pc/MoS$_2$ and MPc/MoS$_2$ (M=Co, Zn) MDHJs are all performed using the Quantum-Espresso package~\cite{Giannozzi2009QE,Giannozzi2017QE} with optimized norm-conserving Vanderbilt pseudopotentials~\cite{Hamann2013} from the PseudoDojo library~\cite{Van2018pseudodojo}. All MDHJs structures are optimized with exchange-correlation potentials using Perdew-Burke-Ernzerhof (PBE) parametrization of the generalized gradient approximation~\cite{Perdew1996}. The PBE-calculated geometries for Pc molecules and 2D MoS$_2$ are within 1\% of experimental results, see Table S1, while previous studies have shown a weak dependence of the electronic structure on the functional used  to optimize the geometries~\cite{Kronik2017CoPc}. The pattern and density of surface organic molecules are derived from  scanning tunneling microscopy~\cite{Liljerothn2014STM}, with an intermolecular distance of $\simeq$ 16 \AA\ (vs 16 \AA\ in experiments~\cite{Liljerothn2014STM}). The corresponding supercells for all MDHJs in this work are 3$\sqrt{3} \times $ 5 in terms of MoS$_2$ unit cell. The plane-wave cutoff energy is 90 Ry. A $k$-point sampling of 2$\times$2 is used for MPc/MoS$_2$ supercells with in-plane size of 16.5$\times$16.0 \AA, or equivalent for other dimensions. The vacuum regions are larger than 21 \AA\ for both monolayer MoS$_2$ and MDHJs. Convergences of total and electronic energy are 10$^{-3}$ eV/atom, and $10^{-6}$ eV, respectively. Thatchenko-Scheffler~\cite{Scheffler2009TSvdw} dispersion corrections are used for van der Waals interactions in MDHJs. Optimally-tuned range-separated hybrid ($\alpha+\beta=1$, see below) and screened range-separated hybrid (SRSH, $\alpha+\beta=1/\varepsilon$, see below) functional calculations for molecules are performed using the NWChem~\cite{Valiev2010nwchem} code, with a cc-PVTZ basis set~\cite{Windus2019Basis} for all atoms. Molecular geometries are optimized with B3LYP~\cite{Lee1988B3LYP}. SRSH calculations for monolayer MoS$_2$ and 0D/2D MDHJs are performed using our customized version of the Quantum-Espresso package that allows for incorporation of the screened exact exchange.

\section{Results and Discussion}
The SRSH functional requires the determination of three (system-dependent) parameters: 1) $\alpha$, the fraction of short-range exact exchange; 2) $\gamma$, the length scale for the short-range to long-range transition; 3) $\beta$ or, precisely, $\alpha+\beta$, the fraction of screened exact exchange in the long range ($\alpha+\beta = 1/\varepsilon$ in order to achieve the correct asymptotic screening of the Coulomb potential at long range~\cite{Kronik2013SRSH,Kronik2015OrgHJ,Kronik2019SRSHGW,ZhengGalli2019}, where $\varepsilon$ is the homogeneous macroscopic dielectric constant). 

\subsection{Electronic Structure of Gas-Phase Phthalocyanines}
\begin{figure*}
	\centering
	\includegraphics[width=0.95\textwidth]{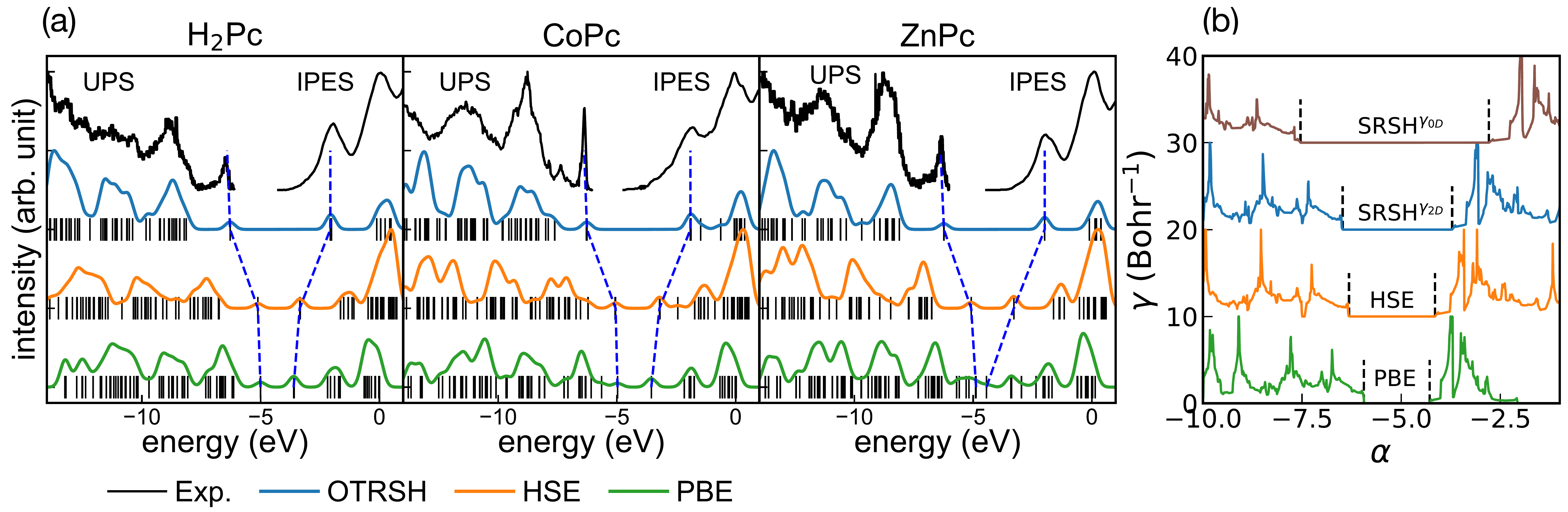}
	\caption{(a) Calculated spectra from optimally-tuned, range-separated hybrid (OT-RSH, cyan) comparing with that from Heyd–Scuseria–Ernzerhof (HSE, orange), Perdew-Burke-Ernzerhof (PBE, green) and experimental UPS~\cite{Berkowitz1979Exp,Ziegler2011UPS} for gas phase and IPES~\cite{Sato2001IPES} for corresponding molecular thin film. A Gaussian broadening of 0.2 eV is used for all spectra from computed energy levels. The highest occupied molecular orbital (HOMO) and  lowest unoccupied molecular orbital  (LUMO) energy levels are connected by the dashed blue lines. Experimental IPES spectra are shifted to align the LUMO peak with that from OT-RSH. (b) Density of states (DOS) for MoS$_2$ from different functionals, PBE, HSE, SRSH with $\alpha=0.1, \alpha+\beta=1$ while $\gamma=0.0245$ Bohr$^{-1}$ (SRSH$^{\gamma_{2D}}$) which is tuned to match GW electronic bandgap of monolayer MoS$_2$, and $\gamma=0.140$ (SRSH$^{\gamma_{0D}}$) as used for Pc molecules, respectively.}%
	\label{fig:dosgas}
\end{figure*} 

In a previous study~\cite{ZhouMPcgas}, we showed that range-separated hybrid functionals obeying Koopmans' theorem~\cite{Baer2009OTRSH,Baer2010OTRSH,Kronik2011OTRSH,Kronik2014OTRSHJCTC,Kronik2012Eg} provide ionization potentials and optical bandgaps in quantitative agreement with experimental results and GW calculations~\cite{Kronik2013CoPcd,Marom2011PRBGW} for MPc molecules in the gas phase (i.e. $\alpha+\beta =1$). In Fig.~\ref{fig:dosgas}, we compare the density of states of the isolated molecules with experimental  ultraviolet photoemission spectroscopy (UPS)~\cite{Berkowitz1979Exp,Ziegler2011UPS} and inverse phototemission spectroscopy (IPES)~\cite{Sato2001IPES}. 
As shown in   Fig.~\ref{fig:dosgas}, the full density of states is in excellent agreement with experiment. Moreover, we note that all parameters  $\alpha, \gamma, \alpha+\beta=1$ yielding a Koopmans' compliant functional (see Fig. S2 and Table S2) provide accurate ionization potentials and electronic affinities~\cite{ZhouMPcgas}. 
In contrast,  $\alpha$ controls the relative energy of non-frontier orbitals associated with $d$ states, with a value of 0.1 yielding a best match with experimental results, see more details in the Supplementary Information, Fig. S2 and Table S2. The optimal parameters are determined to be $\alpha=0.1, \gamma=0.140$ Bohr$^{-1}$ for all the three molecules, which is in good agreement with previous work for gas-phase CoPc molecule~\cite{Kronik2017CoPc}, with the $\simeq4.2$ eV electronic gaps consistent with previous calculations of  H$_2$Pc~\cite{Antonio2017H2PcOTRSH,Kronik2011OTRSH} and CoPc~\cite{Kronik2017CoPc}.

Importantly, predictions using optimally-tuned range-separated hybrids represent a significant improvement over results from PBE and Heyd–Scuseria–Ernzerhof (HSE)~\cite{Heyd2003HSE,Krukau2006HSE} functionals. Although HSE is also a range-separated hybrid functional, it uses universal range-separation parameters that are different from those optimally tuned in the approach of references~\cite{Baer2009OTRSH,Baer2010OTRSH,Kronik2011OTRSH,Kronik2014OTRSHJCTC,Kronik2012Eg}. This highlights the need for a system-dependent tuning of the range-separation parameters.  

\subsection{Electronic Structure of Freestanding MoS$_2$ }

Similarly, optimally-tuned range-separated functionals have been shown to provide band structures in quantitative agreement with GW results for freestanding MoS$_2$~\cite{Ramasubramaniam2019SRSH2D}, in agreement with our results, as shown in Fig. S3 (our optimized values are $\alpha,\gamma=(0.1,0.0245 ~\textrm{Bohr}^{-1})$ as compared with $\alpha, \gamma=(0.106,0.02~ \textrm{Bohr}^{-1})$ of Ref.~\cite{Ramasubramaniam2019SRSH2D}). The density of states obtained with these parameters are shown in Fig.~\ref{fig:dosgas}, and are greatly improved over both PBE and HSE functionals. Importantly, these calculations are carried out for  $\alpha+\beta=1$, in accordance with the fact that the long-range Coulomb interactions between electrons in a 2D material are unscreened~\cite{Qiu2016screen2D,Thygesen2015dielectric2D}. We note that our  range-separated hybrid  functional considers $(\alpha+\beta)$ a constant of $r$, instead of the correct  $r$-dependent dielectric behavior predicted by classical models~\cite{Cho2016MoS2}, resulting in an effective accuracy trade-off between short- and long-range interactions for polarizable 2D materials. 

\subsection{Range Separated Hybrid Functionals for 0D/2D Mixed-Dimensional Heterojunctions}
In contrast to the above, for multicomponent systems such as MDHJs, $\alpha$ and $\gamma$ must be jointly optimized for individual components, while $\alpha+\beta$ must account for the magnitude of the intercomponent Coulomb interactions~\cite{ZhengGalli2019}. This raises several issues as: (1) the optimal set of  $\alpha,\gamma$ for the molecules and the 2D materials found satisfactory above are significantly different (As shown in Fig.~\ref{fig:dosgas}(b), the bandgap of MoS$_2$ is overestimated by 2eV for $\alpha=0.1, \gamma=0.140~ \textrm{Bohr}^{-1}$), (2) long-range screening is anisotropic, with out-of-plane non-local correlation effects decaying asymptotically as $1/\varepsilon r$ due to the polarization of the substrate, while in-plane interactions decay as $1/r$ --although both must be described by isotropic $(\alpha+\beta)/r$ long-range interactions in the SRSH. 

In this work, we solve these challenges by tuning $(\alpha+\beta)/r$ to represent out-of-plane interactions (as explained below), while minimizing the error in the bandgap of freestanding MoS$_2$ for those values of $(\alpha+\beta)$ (as opposed to minimizing the error at $\alpha+\beta=1$).  


\begin{figure*}
	\centering
	\includegraphics[width=0.95\textwidth]{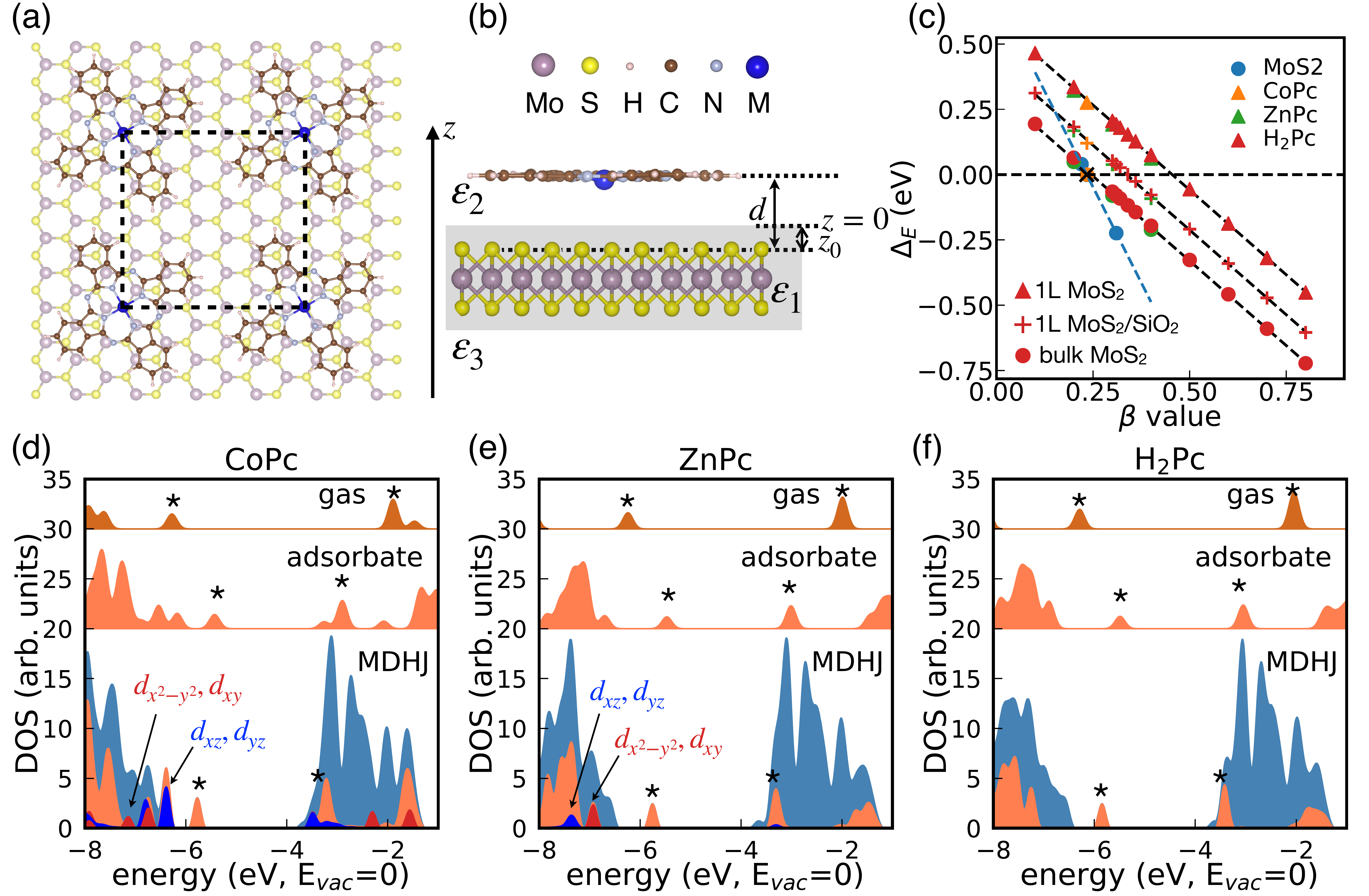}
	\caption{(a) Atomic Structures of phthalocyanine (Pc) molecules on monolayer MoS$_2$. (b) side view of Pc on monolayer MoS$_2$ where $z$=0 is set to be the top surface image plane position, and $z_0, d$ are the normal distance of the surface S sites and Pc molecules, respectively. (c) Changes to the HOMO energy for Pcs and bandgap E$_g$ for MoS$_2$ as a function of $\beta$ values. For Pcs, $\Delta_E$ = HOMO($\alpha=0.1$,$\gamma^{opt}$=0.140 Bohr$^{-1}$,$\beta$) - (HOMO($\alpha=0.1$,$\gamma^{opt}$=0.140 Bohr$^{-1}$,$\beta=0.9$)+$P_{HOMO}$) where $P_{HOMO}$ is the normalization of HOMO from Eq.~\ref{Eq:Pim}. For MoS$_2$, $\Delta_E$=E$_g^{GW}$-E$_g$($\alpha=0.1$,$\gamma=0.140$ Bohr$^{-1}$,$\beta$) where E$_g^{GW}$=2.8 eV is the electronic bandgap from GW calculations~\cite{Cho2016MoS2,Lam2012MoS2scGW,Ram2012MoS2,Yakob2013MoS2,Louie2013MoS2}. The 'x' shows $\beta=0.234$ where the bandgap of ML MoS$_2$ is consistent with the GW result. Data in triangles, '+' and solid dots are results with $P_{HOMO}$ considering different substrates for the molecules.
	(d)-(f) Density of states and projected density of states for gas-phase, gas-phase with $\beta=0.234$ (adsorbate), and MDHJs for CoPc (d), ZnPc (e) and H$_2$Pc (f), respectively. Orange, bright blue, and blue shadowed areas are pDOS for molecules, e$_g (d_{xz},d_{yz})$ orbitals, and monolayer MoS$_2$, respectively. Red shadowed areas are pDOS for b$_{1g} (d_{x^2-y^2})$ and b$_{2g} (d_{xy})$ orbitals. Stars `*'  points to HOMO/LUMO of Pcs. DOS plots for gas-phase Pcs are from eigenvalues with Gaussian broadening of 0.1 eV, while adsorbate refers to isolated Pcs calculated with the same parameters as the MDHJs $\alpha=0.1, \beta=0.5, \gamma=0.05$ Bohr$^{-1}$.}
	\label{fig:ads}
\end{figure*}

Upon molecular adsorption on MoS$_2$ (Fig.~\ref{fig:ads} (a)-(b)), the polarizability of MoS$_2$ stabilizes the charged states of the molecule with respect to its neutral state, leading to, among other effects, a reduction of the molecular bandgap which can be well approximated by a classical electrostatic model~\cite{Neaton2006OrgHJ,Neaton2013HJSigma,Quek2019dielectricHJ}.  Specifically, using $\rho_i (\textbf{r})$ to represent the charge density for molecular state $i$,  the classical energy shift of molecular state $i$ resulting from substrate polarization in the configuration of Fig.~\ref{fig:ads} (a)-(b) is given by Eq.~\ref{Eq:Pim}~\cite{Quek2019dielectricHJ,Kumagai1989dielectric},

\begin{align}\label{Eq:Pim}
	P_i  &= \pm \frac{1}{2} \int_{r}\int_{r^\prime}   \frac{\rho_i (\textbf{r})\rho_i(\textbf{r}^\prime)}{|\textbf{r}-\textbf{r}^\prime|} L_{12} d\textbf{r}d\textbf{r}^\prime \quad \mp \\ \nonumber
	&  \frac{1}{2} \int_{r}\int_{r^{\prime \prime}} \frac{4\varepsilon_1\varepsilon_2}{(\varepsilon_1+\varepsilon_2)^2} L_{13} \sum_{n=0}^{\infty}(L_{12}L_{13})^n \frac{\rho_i (\textbf{r})\rho_i(\textbf{r}_n^{\prime \prime})}{|\textbf{r}-\textbf{r}_n^{\prime \prime}|} d\textbf{r}d\textbf{r}^{\prime \prime}
\end{align}

where $L_{1m}=(\varepsilon_1-\varepsilon_m)/(\varepsilon_1+\varepsilon_m), \textbf{r}=(x,y,z), \textbf{r}^\prime =(x,y,-z) , \textbf{r}_n^{\prime \prime}=(x, y, -z-2(n-1)t)$, with $z, t$ the normal distance from the molecule to the image plane of MoS$_2$, and the thickness of MoS$_2$, respectively. $\varepsilon_1, \varepsilon_2$ and $\varepsilon_3$ are the dielectric constants for the configuration shown in Fig.~\ref{fig:ads}(b).  Monolayer MoS$_2$ is considered as a homogeneous thin dielectric slab with dielectric constant of $\varepsilon=14$~\cite{Lam2012MoS2scGW,Berkelbach2013MoS2}. The first term in Eq.~\ref{Eq:Pim} represents the screening effect of a semi-infinite substrate, while the second term accounts for an infinite series of image charges resulted from two interfaces. The top and side views of the 0D/2D MDHJ atomic structures are shown in Fig.~\ref{fig:ads} (a) and (b), respectively, and are similar for the three molecules (the vertical distances are 3.31 \AA, 3.21 \AA, and 3.25 \AA\ above the top-most sulfur atoms for H$_2$Pc/MoS$_2$, CoPc/MoS$_2$, and ZnPc/MoS$_2$, respectively). We determine that the image plane $z_0$ lies 0.19 \AA\ above the top-most sulfur atoms for MoS$_2$ using the method of reference~\cite{Liu2017OrgHJ,Kronik2015OrgHJ} (see details in the Supplementary Information and Fig. S7). For occupied molecular states, $P_i>0$ while for unoccupied states, $P_i<0$. Similarly, the polarizability of the molecules impacts the bandgap of MoS$_2$. This effect can also described using an electrostatic model ~\cite{Cho2018PRB,Kumagai1989dielectric}, see details in the Supplementary Information. 

In order to tune $\alpha+\beta$ to the correct asymptotic decay of the exchange-correlation potential in the direction normal to the 2D materials, we compute the orbital-dependent energy shifts as a function of $\beta$ for $\alpha= 0.1; \gamma= 0.140\ \textrm{Bohr}^{-1}$, as shown in Fig.~\ref{fig:ads}(c). In agreement with the classical description, the energy shifts in the molecular orbitals as well as in the bandgap of MoS$_2$ are found to vary linearly with $\beta$. 

As shown in Fig.~\ref{fig:ads} (c) and Fig. S5, the magnitude of the classical correction on the HOMO/LUMO energies of the molecules is reached for $\alpha+\beta\simeq 0.6$. However, this value is distinct from the optimal value for the MoS$_2$ bandgap ($\alpha+\beta\simeq 0.33$) and a joint minimization of these errors as a function of  ($\alpha, \gamma$)  is necessary, as shown in Fig. S4. Moreover, Eq.~\ref{Eq:Pim} allows to consider the effects of different dielectric environments on the non-local correlations through the tuning of the thickness and $\varepsilon_3$ i.e. 1L MoS$_2$, bulk MoS$_2$ as well as 1L MoS$_2$ on a bulk SiO$_2$ substrate (assuming $\varepsilon_{SiO_2}=3.9$~\cite{Robertson2004SiO2eps}), leading to different $P_{HOMO}$, see Fig.~\ref{fig:ads} (c).

For the case of phthalocyanine molecules on 1L MoS$_2$, we determine the optimized SRSH functional parameters to be $\alpha=0.1, \beta=0.5, \gamma=0.05$ Bohr$^{-1}$, with an estimated total error in the bandgap of MoS$_2$ and orbital energies of phthalocyanine molecules about 0.25 eV, see details in Fig. S4 of the Supplementary Information. While we tune the RSH parameters considering only the HOMO energies here, we note that the changes for LUMO energy are consistent with that for the HOMO, as shown in Fig. S5, though there are slight variations among different phthalocyanine molecules. As shown above, SRSH functionals can be optimally tuned  for any MDHJs using the procedure above. In general cases, it may require tuning of $\alpha, \beta$ together in order to both accurately describe the ($\alpha+\beta$)/r long-range interactions and minimize the band gap error for MoS$_2$.

Importantly, in our MDHJ calculations, $\alpha+\beta \neq 1$, different from freestanding MoS$_2$. Density of states for monolayer MoS$_2$ with the parameters of the MDHJ ($\alpha=0.1, \beta=0.234, \gamma=0.140$ Bohr$^{-1}$), and freestanding MoS$_2$ $\alpha+\beta= 1$ and $\alpha=0.1, \gamma=0.0245$ Bohr$^{-1}$  are shown in Fig. S6, and show negligible differences.

\subsection{Electronic Structure of Phthalocyanines on MoS$_2$}

In Fig.~\ref{fig:ads} (d-f), we show the projected density of states (pDOS) of H$_2$Pc/MoS$_2$, CoPc/MoS$_2$ and ZnPc/MoS$_2$ MDHJs using our optimized SRSH functional with $\alpha=0.1,\beta=0.5,\gamma=0.05$ Bohr$^{-1}$. 

The optimized SRSH functional leads to a decrease in the molecular HOMO-LUMO gap of about 1.9 eV  
upon absorption as shown in Fig.~\ref{fig:ads} (d-f), due to dielectric screening from MoS$_2$.  The HOMO-LUMO gaps of the gas-phase molecules are about 4.2 eV, and fall to 
2.20 eV, 2.36 eV and 2.35 eV respectively for CoPc, ZnPc, and H$_2$Pc on MoS$_2$, in very good agreement with experimental UPS and IPES results of 2.2 eV for H$_2$Pc in the H$_2$Pc/MoS$_2$ MDHJ~\cite{Mutz2020PcMoS2}. 

All 3 MDHJs are found to have a type-II band alignment, with the HOMO of Pcs within the band gap of MoS$_2$, and LUMOs within the conduction band of MoS$_2$. In contrast, the $d$-states of  ZnPc and CoPc associated with the HOMO-1 and HOMO-2 are more impacted by the dielectric screening than the HOMO/LUMO due to their higher localization, a behavior captured by Eq.\ref{Eq:Pim} and by gas-phase calculations at $\alpha + \beta < 1$ . Upon adsorption, we observe an interface charge redistribution with the molecules donating a fraction of an electron to MoS$_2$, leading to downward shift of $\simeq 0.3$ eV for all molecular orbitals. This fractional charge transfer can be attributed to the filling of induced density of interface states in the band gap of monolayer MoS$_2$~\cite{Flores2006CT,Flores2008CT,Kahn2006CT}. The ground-state charge density change have been shown in Fig. S6 of previous work~\cite{Amsterdam2019PcMDHJ}. We denote this near-uniform shift of the molecular orbitals as $\Delta E_{FO}$ and report its value in Table~\ref{tbl:Sigma}. 

 Interestingly, these different effects --dielectric screening effect and interface dipole, result in the doubly-degenerate HOMO-1 ($d_{yz},d_{xz}$) of CoPc  at near-resonance with the valence band maximum (VBM) of MoS$_2$ suggesting a possible resonant coupling with MoS$_{2}$ holes. This is in contrast from ZnPc in which the HOMO-1 is at lower energy and has a $d_{x^2-y^2},d_{xy}$ character. This suggests that the experimental observations of metal-dependent electronic properties in MPc-based MDHJs~\cite{Amsterdam2019PcMDHJ} could result from the resonance of the metal $d$-states of out-of-plane $d$-character with the VBM of MoS$_2$. As shown in  Fig. S10, these findings are in contrast with the PBE predictions for CoPc/MoS$_2$, ZnPc/MoS$_2$, and H$_2$Pc/MoS$_2$ MDHJs. 

\subsection{Model Self-Energy Corrections}

We note that the electronic structure of the gas-phase molecules in the dielectric background ($\beta=0.5$) in Fig.~\ref{fig:ads} (d-f) is highly reminiscent of that of the MDHJs, suggesting the possibility of predicting the electronic structure of MDHJs from a perturbative correction to gas-phase properties and interface charge redistribution~\cite{Neaton2006OrgHJ,Neaton2009adsSigmaa}, which we now derive. 

\begin{figure*}
	\centering
	\includegraphics[width=0.9\textwidth]{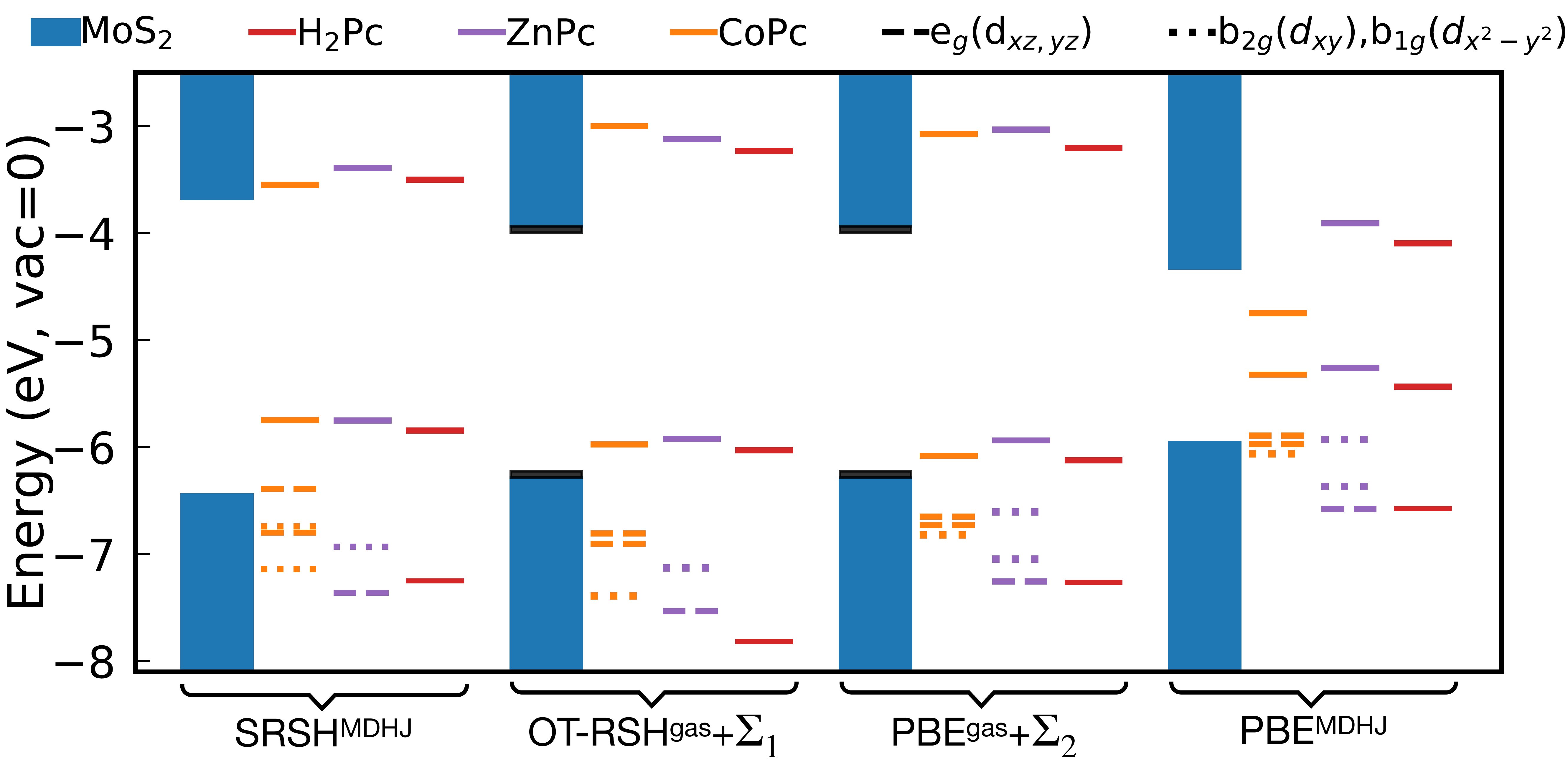}
	\caption{Energy levels for MoS$_2$ (blue), CoPc (orange), ZnPc (purple), and H$_2$Pc (red) at Pc/MoS$_2$ MDHJ from SRSH$^{MDHJ}$ for MDHJ, OT-RSH$^{gas}+\Sigma_1$ ($\Sigma_1$=$P$+$\Delta E_{FO}^{PBE}$ to add dielectric screening, Eq.~\ref{Eq:Pim}, and interface dipole effects, respectively, to the gas-phase OT-RSH energy levels), PBE$^{gas}+\Sigma_2$ ($\Sigma_2$ includes $\Sigma_1$ as well as adding corrections to the PBE results using $\Delta$SCF method~\cite{Jones1989DFT,Neaton2009adsSigmaa,Darancet2012NL}) , and PBE results for MDHJ. $\Delta E_{FO}^{PBE}$ is near-rigid shift of band edges for isolated components upon forming heterojunctions. Those values are included in Table~\ref{tbl:Sigma}. HOMO/LUMO for Pcs are in solid lines and orbital energies below HOMO are shown as dashed lines for e$_g(d_{xz},d_{yz})$ 
		orbitals and dotted lines for b$_{2g} (d_{xy})$ and b$_{1g} (d_{x^2-y^2})$ orbitals. For MoS$_2$, $\Sigma$ accounts for the dielectric confinement~\cite{Kumagai1989dielectric,Cho2018PRB} where the black shadowed areas depicts the increase/decrease of VBM/CBM after adsorption of Pc, see detailed method and parameters in the Supplementary Information.  
	}
	\label{fig3:edges}
\end{figure*} 

Following Neaton et al.~\cite{Neaton2006OrgHJ,Neaton2009adsSigmaa}, we compute the self-energy corrections to the molecular levels in the gas phase and to the band edges of MoS$_2$ resulting from non-local correlations and interface charge distributions, the former $P$ from the classical electrostatic model above and the latter, $\Delta E_{FO}$, from a DFT calculation of the MDHJ. Importantly, we note that the shift caused by interface charge redistribution predicted by the PBE functional is within 
0.11 eV of the one predicted by the SRSH functional (Table \ref{tbl:Sigma}), implying this effect can be captured by lower-level calculations. Unlike Koopmans' compliant range separated hybrids, PBE also needs to be corrected for the error in the description of the individual components. Past works~\cite{Neaton2009adsSigmaa,Darancet2012NL} have used a scissor operator based on $\Delta$SCF~\cite{Jones1989DFT} method for organic molecules, though limits of this approach have been pointed out for organometallic complexes~\cite{LiuPorphyrins2014} suggesting the use of hybrid functional as a starting point for this system.  These scissor corrections to the PBE-calculated HOMO/LUMO energies, denoted as $\Sigma_{GP}$, are included in Table~\ref{tbl:Sigma} for all three molecules. The resulting HOMO/LUMO gaps are around $4.2 \sim 4.3$ eV, close to that of 4.2 eV from the range-separated hybrid functional for H$_2$Pc, CoPc, and ZnPc. For MoS$_2$, we apply a model self-energy correction to the PBE band edges following Refs. \cite{Kumagai1989dielectric,Cho2018PRB}.

\begin{table*}
	\small
	\captionsetup{margin={1.0cm,1.0cm}}
	\caption{\ Magnitude of each correction for HOMO (LUMO)/valence (conduction) band energies as discussed in Fig.~\ref{fig3:edges}. All energy values are in eV. Using the charge densities of HOMO-1 state, $P$ values for HOMO-1 orbitals are 0.50 eV, 1.06 eV, 1.16 eV for H$_2$Pc, ZnPc and CoPc, respectively.}
	\label{tbl:Sigma}
	\begin{tabular*}{0.7\textwidth}{@{\extracolsep{\fill}}cccccc}
		\hline
		Pc@HJ & $\Sigma_{GP}$ & $P$ & $\Delta E_{FO}$(OT-SRSH) & $\Delta E_{FO}$(PBE) \\ \hline
		CoPc & -1.35 (1.43) &   0.60 (0.78) & -0.34 (-0.30) &  -0.45 (-0.31) \\ 
		ZnPc & -1.26 (1.62) &   0.58 (-0.74) & -0.28 (-0.38) &  -0.29 (-0.36) \\ 
		H$_2$Pc & -1.26 (1.62) &  0.57 (-0.73) & \-0.35 (-0.44) &  -0.43 (-0.48) \\ 
		MoS$_2$ &  -0.59(0.55) & $\sim$0.1 & $<$0.06 & $<$0.05 \\ 
		\hline
	\end{tabular*} 
\end{table*}

The resulting self-energy corrections to the molecular levels are denoted OT-RSH$^{gas}+\Sigma_1$ where $\Sigma_1$=$P$+$\Delta E_{FO}^{PBE}$, and PBE$^{gas}+\Sigma_2$ where $\Sigma_2 = \Sigma_1 + \Sigma_{GP}$ and included in Fig.~\ref{fig3:edges}. We note that the model of Ref. \cite{Cho2018PRB} is also able to predict the dielectric screening effect of the Pc molecules on the energy levels of MoS$_2$, as shown in Fig.~\ref{fig3:edges} by the black region, an increase(decrease) on the VBM(CBM) by about 0.1 eV (calculation details are included in the Supplementary Information). Similarly, we can account for the dielectric screening effect of the SiO$_2$ substrate and H$_2$Pc on MoS$_2$. By using dielectric constants for MoS$_2$, SiO$_2$ and H$_2$Pc of 14~\cite{Lam2012MoS2scGW,Berkelbach2013MoS2}, 3.9~\cite{Robertson2004SiO2eps}, and 1.9~\cite{Shi2007PcDielectric}, respectively, we obtain a bandgap of 
2.3 eV for 1L MoS$_2$ in good agreement with the experimental values of 2.1 eV~\cite{Mutz2020PcMoS2} for monolayer MoS$_2$ in H$_2$Pc/MoS$_2$ on SiO$_2$ substrate.  According to Eq.~\ref{Eq:Pim}, we are able to account for the effect of different dielectric background on the orbital energies of the molecules, e.g. 1L MoS$_2$, bulk MoS$_2$ and 1L MoS$_2$/SiO$_2$. The energy levels for the molecules and 1L MoS$_2$ at those different dielectric backgrounds are included in Fig. S8 in the Supplementary Information.

The level alignment predicted by these self-energy corrections as compared with PBE and SRSH calculations for the three MDHJs is shown in Fig.~\ref{fig3:edges}.  Both self-energy corrections predict HOMO, LUMO, VBM, CBM energies in close agreement with the SRSH calculations and represent a significant improvement over the PBE results. In contrast, and in agreement with the conclusions of Liu et al.~\cite{LiuPorphyrins2014}, the HOMO-1, HOMO-2, and HOMO-3 levels associated with the $d$-orbitals of CuP, CoP molecules are not accurately corrected by the scissor operator in the PBE$^{gas}+\Sigma_2$ approach, which provide non-frontier 3$d$-orbitals for ZnPc 
0.33 eV above the OT-RSH$^{gas}+\Sigma_1$, as shown in Fig.~\ref{fig3:edges}. As a result, the PBE$^{gas}+\Sigma_2$ approach results in qualitative error in the position of the non-frontier orbitals with respect to the VBM energy of MoS$_2$. In contrast, the OT-RSH$^{gas}+\Sigma_1$ approach does not suffer from this deficiency, indicating the importance of an accurate representation of the $d$-orbitals in the gas-phase calculations. 

While SRSH functionals for MDHJs can produce results comparable to experiments as shown above, this model circumvent the need for joint optimization of the parameters in the SRSH functional, as it requires only gas-phase calculations and standard PBE calculations for the 2D and heterojunction structures, while accounting for the different dielectric environments.

\section{Summary}
In summary, we have obtained the electronic structures with electronic bandgaps in quantitative agreement with photoemission experiments using optimally-tuned, screened range-separated hybrid functionals for the 0D/2D H$_2$Pc/MoS$_2$, CoPc/MoS$_2$, and ZnPc/MoS$_2$ mixed-dimensional heterojunctions. We showed that the molecules are strongly impacted by the dielectric environment of the MDHJs, reducing their bandgap by about 2.0 eV for MPc molecules on MoS$_2$. The HOMO-1 non-frontier orbital has large $d$ orbital contribution and overlap with the VBM of MoS$_2$ for CoPc, and is pushed down in energy for ZnPc. This metal-dependent behavior of MPc-based MDHJs can be accurately captured by model, parameter-free corrections to gas-phase calculations, demonstrating that this surrogate can help guide the design and realization of organic-inorganic mixed-dimensional 0D/2D heterojunctions.

\section*{SUPPLEMENTARY MATERIAL}
See supplementary material for the electronic structure of H$_2$Pc, CoPc and ZnPc from $\alpha$ of 0.1, 0.2, and 0.3, comparing with experimental UPS and IPES spectra; DOS and pDOS for the 0D/2D MDHJs from PBE calculations;  Details of the electrostatic model accounting for both self-interaction energy correction to PBE results, as well as non-local dielectric screening effect. Molecular structures in XYZ format, and MDHJ structures in crystallographic information file (cif) format. \\

\section*{ACKNOWLEDGEMENTS}
This work was supported by the Northwestern University MRSEC under National Science Foundation grant No.~DMR-1720139 (Q.Z, P.D., and T.J.M). Use of the Center for Nanoscale Materials (CNM), an Office of Science user facility, was supported by the U.S. Department of Energy, Office of Science, Office of Basic Energy Sciences, under Contract No. DE-AC02-06CH11357. We gratefully acknowledge use of the Bebop cluster in the Laboratory Computing Resource Center at Argonne National Laboratory. We are indebted to an anonymous reviewer for noticing a mistake in our original implementation of Eq.~\ref{Eq:Pim}.

\section*{CONFLICT OF INTEREST}
The authors have no conflicts to disclose.

\section*{DATA AVAILABILITY}
The data that support the findings of this study are available from the corresponding author upon reasonable request.

%

\end{document}


\preprint{}


\title{SUPPORTING INFORMATION\\ \vspace{15pt}
	Range-Separated Hybrid Functionals for Mixed Dimensional Heterojunctions: Application to Phthalocyanines/MoS$_2$}

\author{Qunfei~Zhou}
\affiliation{Materials Research Science and Engineering Center, Northwestern University, Evanston, IL 60208, USA}
\altaffiliation{Center for Nanoscale Materials, Argonne National Laboratory, Argonne, IL 60439, USA}

\author{Zhen-Fei Liu}
\email{zfliu@wayne.edu}
\affiliation{Department of Chemistry, Wayne State University, Detroit, MI 48202, USA}

\author{Tobin J. Marks}
\email{t-marks@northwestern.edu}
\affiliation{Department of Chemistry and Department of Materials Science and Engineering, Northwestern University, Evanston, IL 60208, USA}
\altaffiliation{Materials Research Science and Engineering Center, Northwestern University, Evanston, IL 60208, USA}

\author{Pierre~Darancet}
\email{pdarancet@anl.gov}
\affiliation{Center for Nanoscale Materials, Argonne National Laboratory, Argonne, IL 60439, USA}
\altaffiliation{Northwestern Argonne Institute for Science and Engineering, Evanston, IL 60208, USA}

\maketitle
\raggedbottom

\section{Geometries}

\begin{figure} [H]
	\captionsetup[subfigure]{justification=centering}
	\centering
	\includegraphics[width=0.7\textwidth]{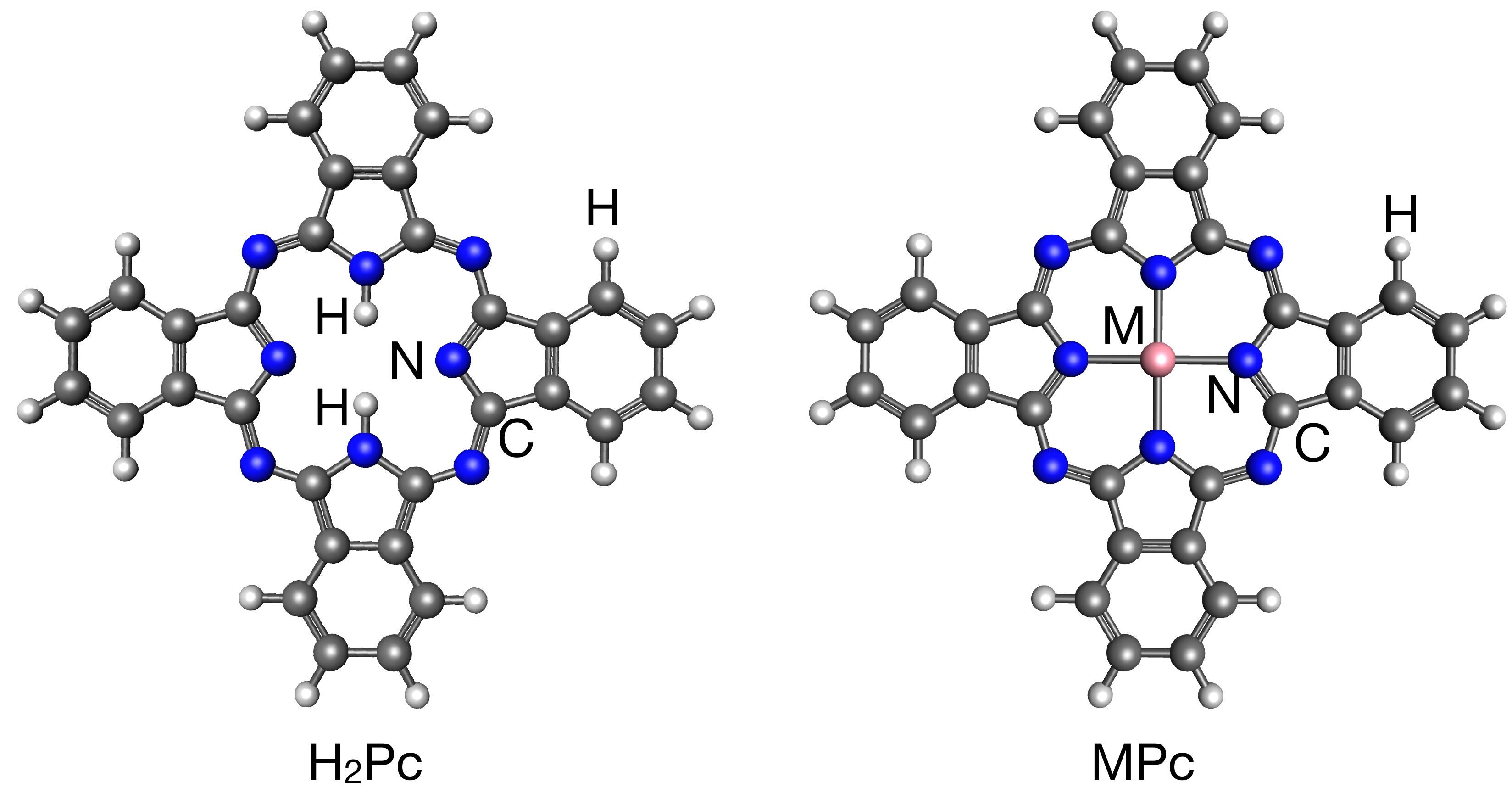}
	\caption{Molecular structure of H$_2$Pc and MPc (M=Co, Zn). }
	\label{fgr:Pcstr}
\end{figure}

\begin{table}[H]

	\small
	\centering
	\caption{Computed structural parameters for MPc molecules. R$_{M-N}$ and R$_{N-C}$ are the bond lengths of metal-nitrogen atoms, and nitrogen-carbon atoms, respectively. The atoms are labeled in Fig.~\ref{fgr:Pcstr}. } 
	\label{tbl:strs}
	\begin{tabular*}{0.8\textwidth}{@{\extracolsep{\fill}}lcccccc}
		\hline
		\multirow{2}{*}{MPc} & \multicolumn{3}{c}{R$_{M-N}$ (\AA)} & \multicolumn{3}{c}{R$_{N-C}$ (\AA)} \\ \cline{2-4} \cline{5-7}
		
		 & B3LYP & PBE & exp.$^a$ &  B3LYP & PBE & exp.$^a$ \\ \hline
		CoPc	&1.936	& 1.919 & 1.908-1.915  &  1.373 & 1.387 & 1.382 \\ 
		ZnPc 	& 1.999 & 1.998 & 1.980 & 1.367 & 1.376 & 1.374 \\ 
		H$_2$Pc & - & - & - & 1.373 & 1.370 & 1.370 \\ 
		\hline
		& \multicolumn{4}{c}{lattice constant a (\AA)} & & \\ \cline{2-5}
		& LDA~\cite{Ramasubramaniam2019SRSH2D} & PBE~\cite{Reed2014MoS2} & PBE (this work)& exp.~\cite{Seifert2007MoS2} & & \\\cline{2-5}
		MoS$_2$ & 3.16 & 3.18 & 3.18 & 3.16 & & \\
		\hline
	\end{tabular*} 
	\\
	\parbox{0.9\textwidth}{
		$^a$  CoPc from neutron-diffraction~\cite{Williams1980CoPc}; ZnPc from XRD~\cite{Scheidt1977ZnPc}, H$_2$Pc from XRD and neutron-diffraction~\cite{Hamer1997H2PcExp}%
	}
\end{table}

\section{DOS for gas-phase molecules}
\addcontentsline{toc}{section}{DOS for gas-phase molecules}

\begin{figure} [H]
	\captionsetup[subfigure]{justification=centering}
	\centering
	\includegraphics[width=0.9\textwidth]{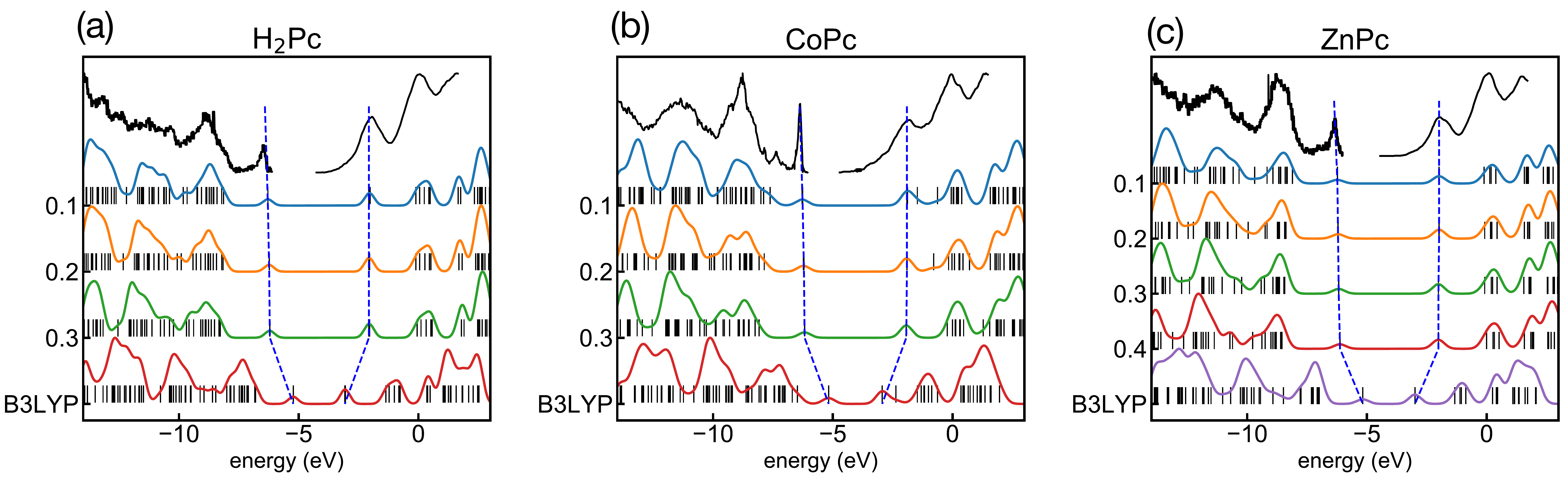}
	\caption{Calculated spectra of H$_2$Pc, CoPc, and ZnPc from OT-RSH with different $\alpha$ values shown as $y$-axis labels and $\gamma^{opt}$ in Table~\ref{tbl:gamma} for each $\alpha$, and comparing with experimental UPS and IPES results (black) lines. HOMO and LUMO changes slightly with $\alpha$, and results from $\alpha=0.1$ has best agreement with experiments. Spectra are all plotted from computed eigenvalues with a Gaussian broadening of 0.2.  Experimental IPES spectra are shifted to align the LUMO peak with that from OT-RSH with $\alpha=0.1$. }
	\label{fgr:alphas}
\end{figure}

Here we compare the electronic structure of gas-phase H$_2$Pc, CoPc and ZnPc molecules from OT-RSH with $\alpha$ of values 0.1, 0.2, 0.3 with experimental UPS~\cite{Berkowitz1979Exp,Ziegler2011UPS} and IPES~\cite{Sato2001IPES} results, see Fig.~\ref{fgr:alphas}. For each $\alpha$, corresponding $\gamma$ are optimally tuned based on the Koopman's theorem, and values are included in Table~\ref{tbl:gamma}. Details on the theory and results can be found in our previous work~\cite{ZhouMPcgas}. As shown in Fig.~\ref{fgr:alphas}, the non-frontier orbitals (i.e HOMO-1, LUMO+1) shift slightly with different $\alpha$ values and the spectra from $\alpha$=0.1 has best agreement with experiments. As a result, we determine the optimal parameters as $\alpha=0.1$, $\gamma$ values in Bohr$^{-1}$ for H$_2$Pc, CoPc and ZnPc are 0.141, 0.140 and 0.139 respectively for gas-phase molecules. As the orbital energies change negligibly ($<$ 0.01 eV) with the three $\gamma$ values, we use $\alpha$=0.1, $\gamma$=0.140 Bohr$^{-1}$ for all SRSH calculations of MDHJs. 

\begin{table} [H]
	\small
	\caption{\ Optimally-tuned $\gamma^{opt}$ (Bohr$^{-1}$) for $\alpha$ of 0.1, 0.2, 0.3, and corresponding HOMO and LUMO energies for gas-phase H$_2$Pc, CoPc, and ZnPc molecules.} 
	\label{tbl:gamma}
	\begin{tabular*}{0.95\textwidth}{@{\extracolsep{\fill}}llllllllll}
		\hline
		\multirow{2}{*}{$\alpha$}  & \multicolumn{3}{c}{H$_2$Pc} & \multicolumn{3}{c}{CoPc} & \multicolumn{3}{c}{ZnPc} 
		\\ \cline{2-4} \cline{5-7} \cline{8-10}
		& $\gamma^{opt}$ & HOMO & LUMO  & $\gamma^{opt}$ & HOMO & LUMO & $\gamma^{opt}$ & HOMO & LUMO\\  
		\hline
		0.1 & 0.141 & -6.286 & -2.068 & 0.140 & -6.258  & -1.905 & 0.139 & -6.231 & -1.986    \\ 
		0.2	& 0.121 & -6.231 & -2.068 & 0.123 & -6.204  & -1.905 & 0.122 & -6.204 & -1.986 \\  
		0.3 & 0.104 & -6.204 & -2.068 & 0.102 & -6.177  & -1.932 & 0.100 & -6.177 & -2.014  \\ 
		\hline
	\end{tabular*} 
\end{table}		

\section{Tuning Range-Seperated Hybrid Functionals for 2D MoS$_2$ and Adsorbed Molecules}
\addcontentsline{toc}{section}{SRSH functional tuning for monolayer MoS$_2$}

 \begin{figure}[H] 
	\centering
	\includegraphics[width=0.7\textwidth]{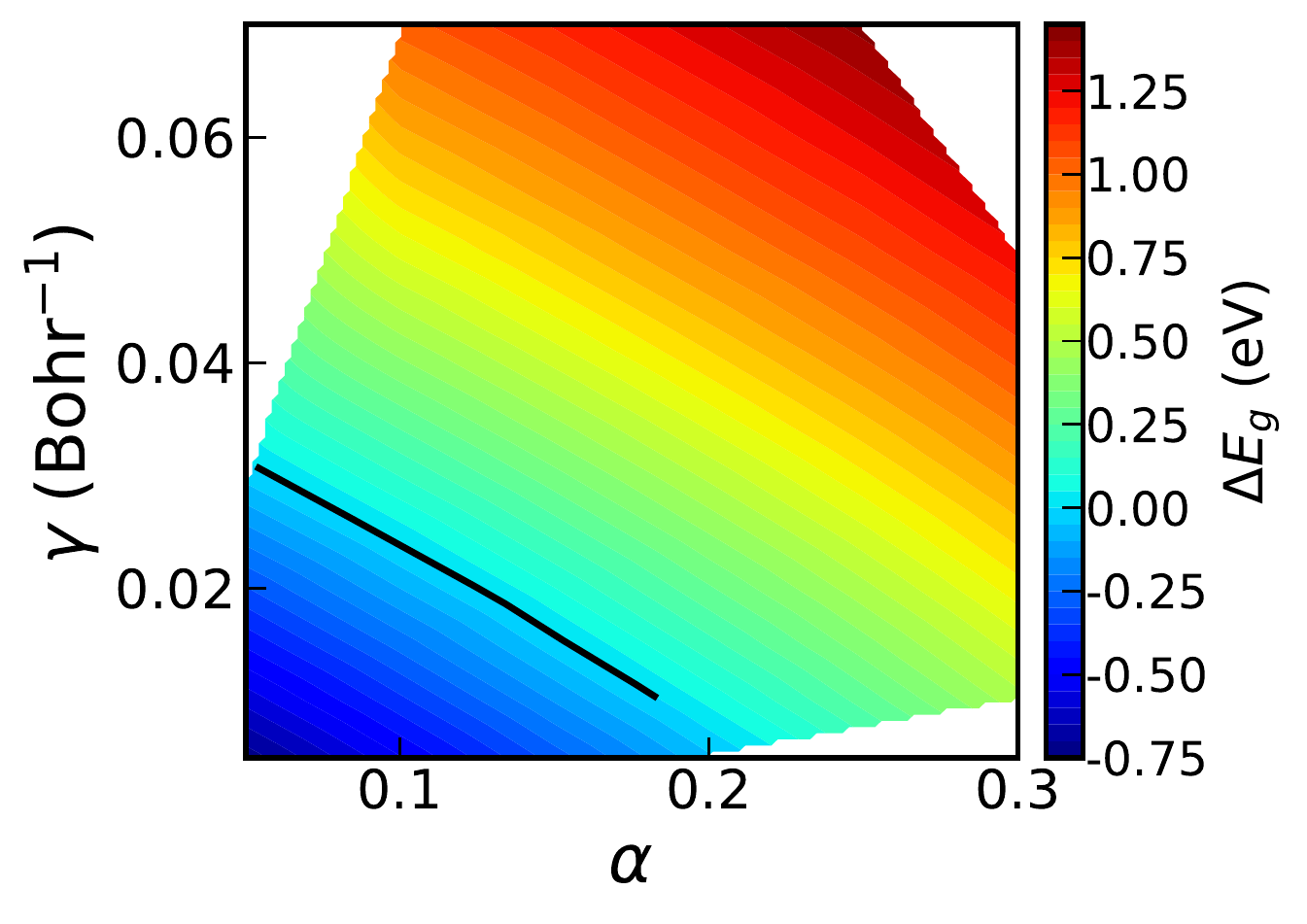}
	\caption{Difference in electronic band gap $\Delta$E$_g$ between SRSH and GW (E$_g^{GW}$=2.8 eV)~\cite{Cho2016MoS2,Lam2012MoS2scGW,Yakob2013MoS2,Louie2013MoS2} calculations, as a function of $\alpha$ and $\gamma$. $\beta=1/\varepsilon_\infty - \alpha$ where $\varepsilon_\infty=1$ for 2D MoS$_2$ layer in order to achieve the correct asymptotic screening of the Coulomb potential~\cite{Ramasubramaniam2019SRSH2D}. The solid black line indicates where the E$_g$ from SRSH equals to E$_g^{GW}$.} 
	\label{figS:2DRSHTune}
\end{figure}

 \begin{figure}[H] 
	\centering
	\includegraphics[width=0.8\textwidth]{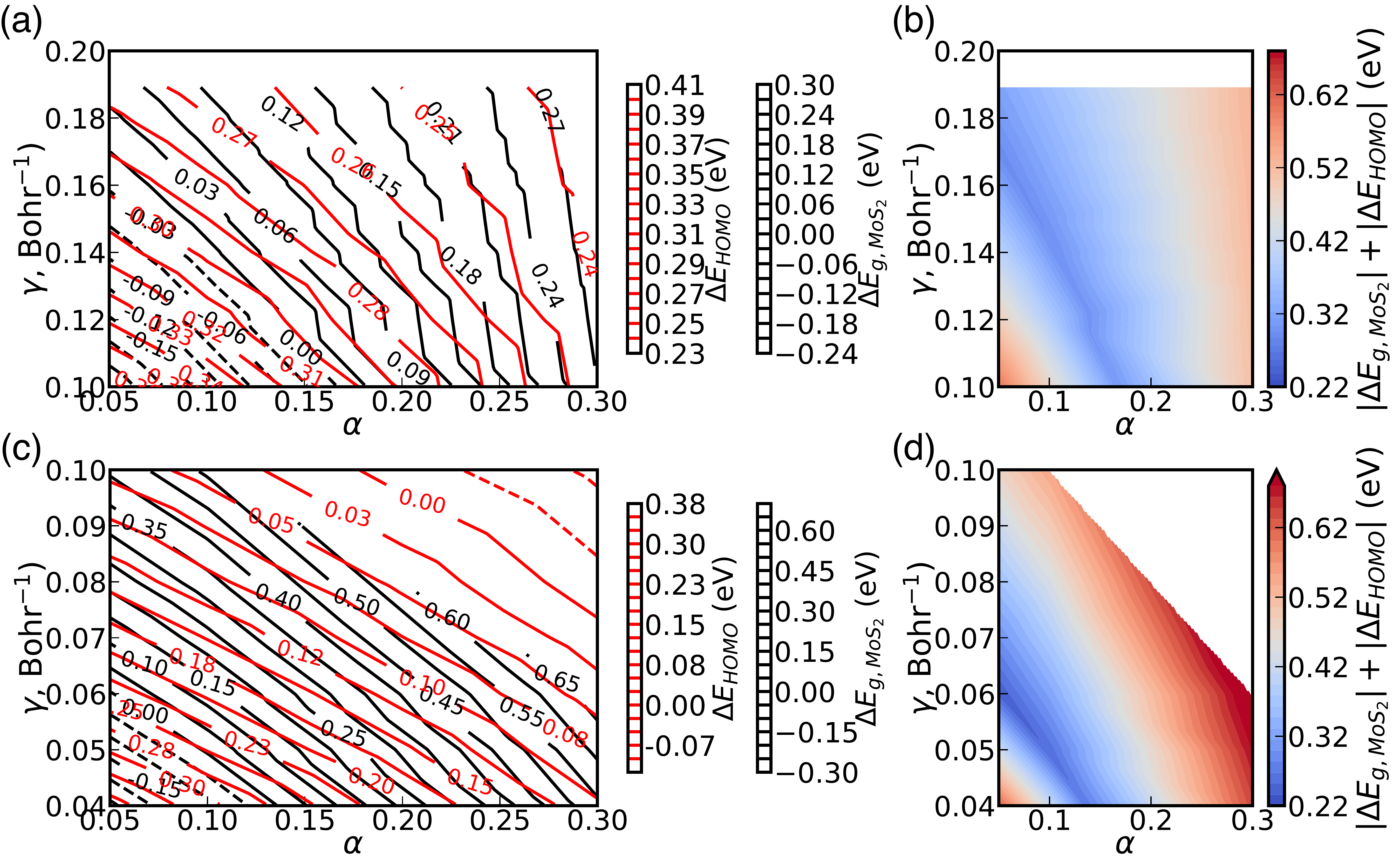}
	\caption{Changes to the HOMO energy for phthalocyanine and bandgap E$_g$ for MoS$_2$ as a function of $\alpha, \gamma$ values for $\alpha+beta=0.33$ (a-b) and $\alpha+\beta=0.60$ (c-d). $\Delta E_{HOMO} = E(\alpha,\beta, \gamma) - (E_{HOMO} + P_{HOMO})$ where $P_{HOMO}$ is from Eq. 1 for molecules on 1L MoS$_2$. $\Delta E_{g,MoS_2} = E_{MoS_2} - E_g^{GW}$ where $E_g^{GW}$=2.8~\cite{Cho2016MoS2}. Based on the error for the band gap of 1L MoS$_2$ and energy level of phthalocyanine as shown in (b) and (d), we choose RSH parameters of $\alpha=0.1, \beta=0.5, \gamma=0.05$ Bohr$^{-1}$ where $|\Delta_{g,MoS_2}| + |\Delta E_{HOMO}|$ is about 0.25 eV.}
	\label{figS:dEAB}
\end{figure}

 \begin{figure}[H] 
	\centering
	\includegraphics[width=0.5\textwidth]{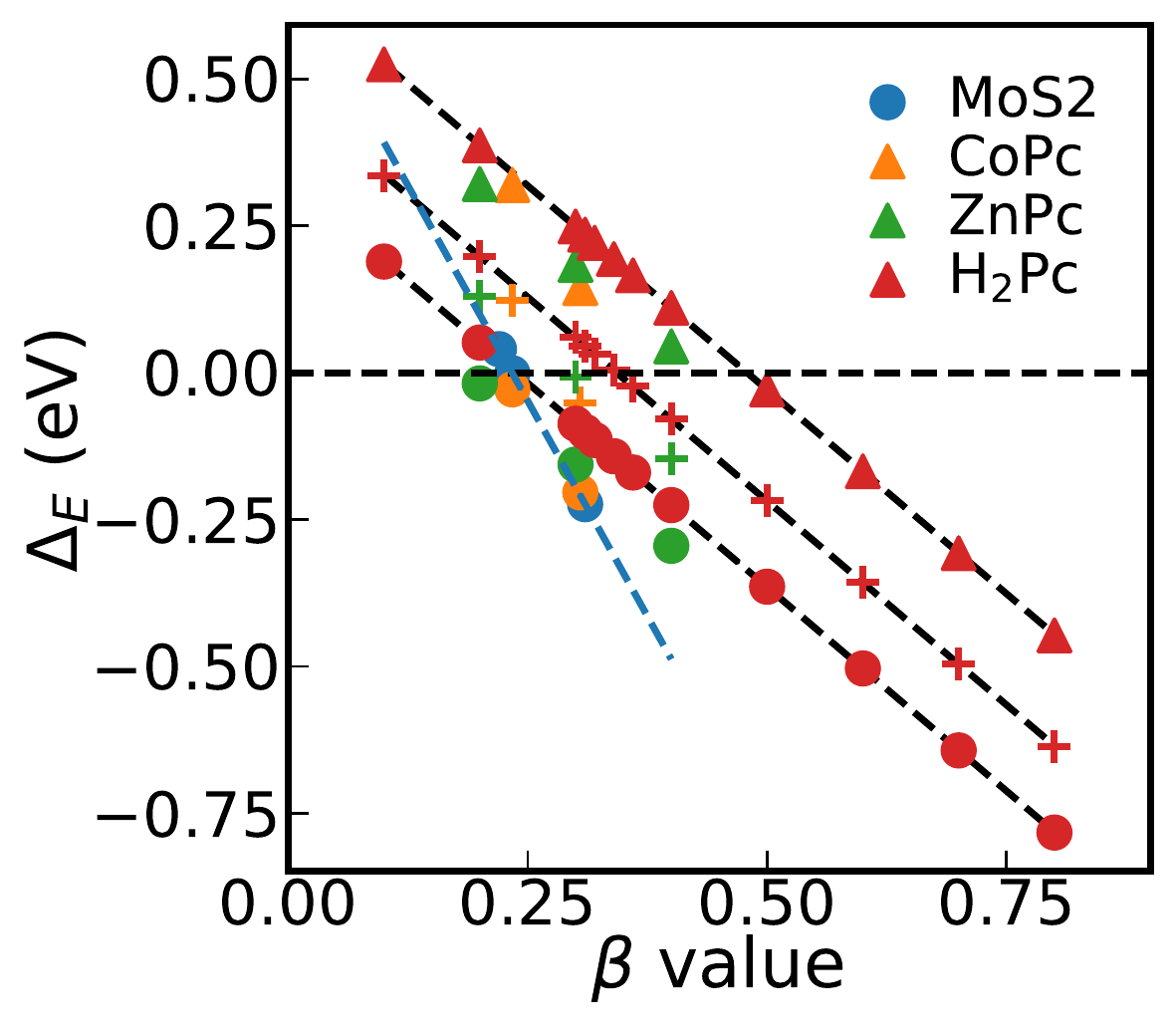}
	\caption{Same as in Fig. 2(c), while $\Delta_{E}$ for the molecules are for LUMO energies.}
	\label{figS:dElumobeta}
\end{figure}


 \begin{figure}[H] 
	\centering
	\subfloat[]{\includegraphics[width=0.4\textwidth]{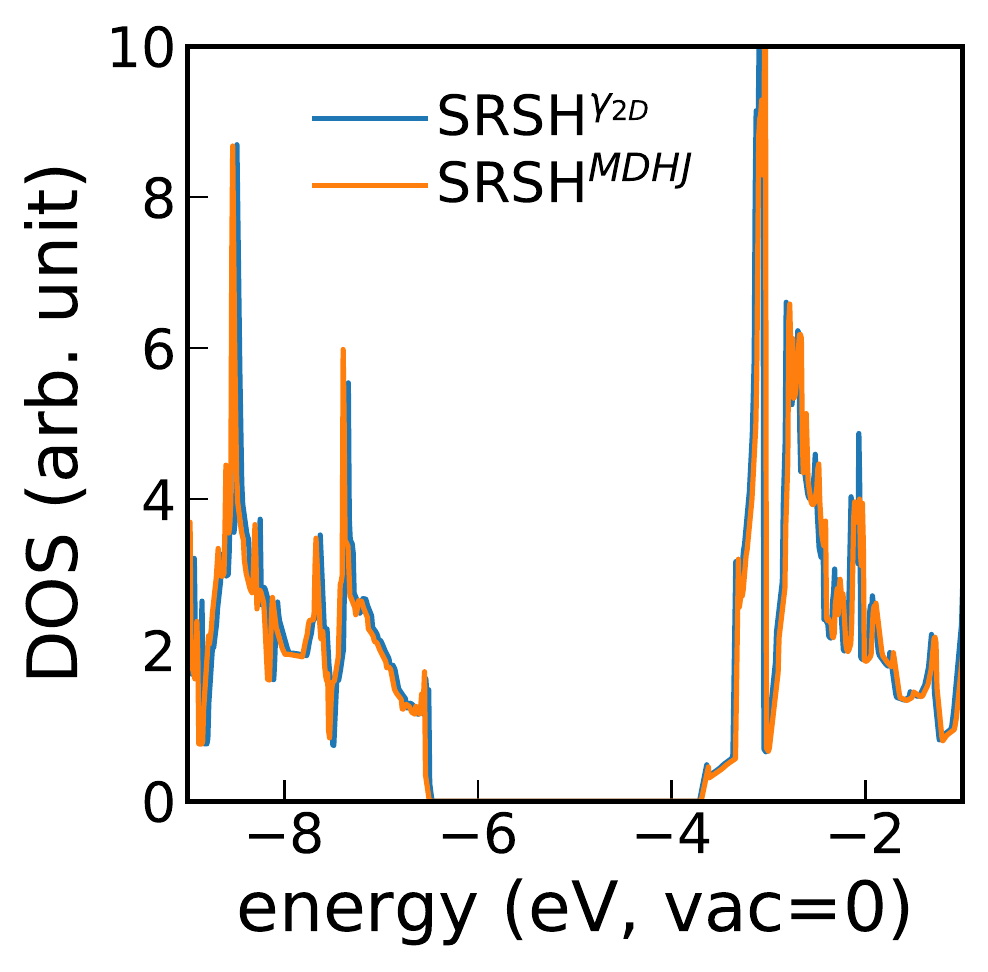}} \hspace{12pt}
	\subfloat[]{\includegraphics[width=0.4\textwidth]{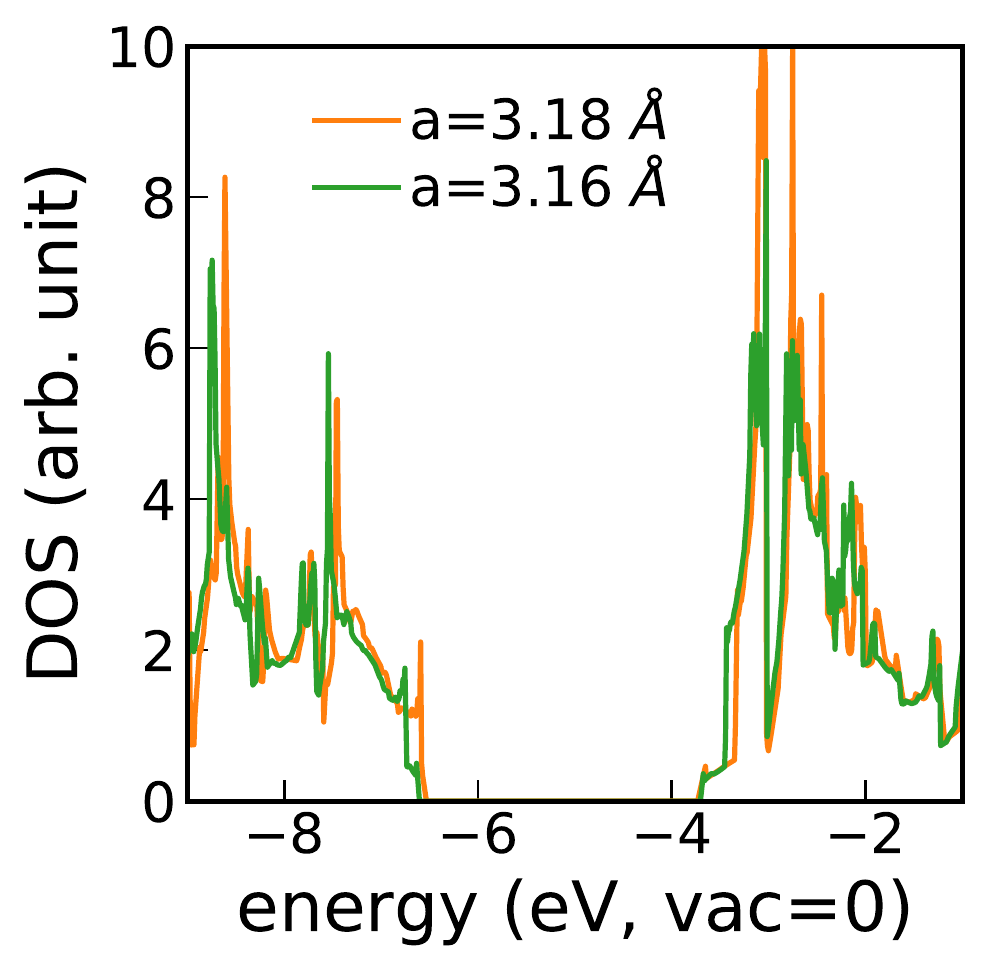}}
	\caption{(a) Density of states for monolayer MoS$_2$ from SRSH functionals with $\alpha=0.1,\gamma=0.0245$ Bohr$^{-1}$ (SRSH$^{\gamma_{2D}}$) where $\alpha+\beta=1$, and $\alpha=0.1,\beta=0.5,\gamma=0.05$ Bohr$^{-1}$ (SRSH$^{MDHJ}$) which are determined for the MDHJs. 
	 (b) DOS of monolayer MoS$_2$ from SRSH$^{MDHJ}$ functionals with lattice constant $a$ of 3.18 \AA\ and 3.16 \AA. The DOS changes negaligibly with lattice parameter.}
	\label{figS:dosMoS2beta}
\end{figure}

As shown in Fig.~\ref{figS:dosMoS2beta}, the DOS of monolayer MoS2 from two SRSH functionals describing different long-range Coulomb screening, $\alpha+\beta$ of 1 and 0.334, respectively, as well as distinct range-separation parameter $\gamma$, 0.0245 Bohr$^{-1}$ and 0.140 Bohr$^{-1}$, respectively. The corresponding length scales for the onset of the long-range Coulomb screening are about 21.60 and 3.78~\AA, respectively.

\section{Image potential model}

\begin{figure}[H]
	\centering
	\includegraphics[width=0.4\textwidth]{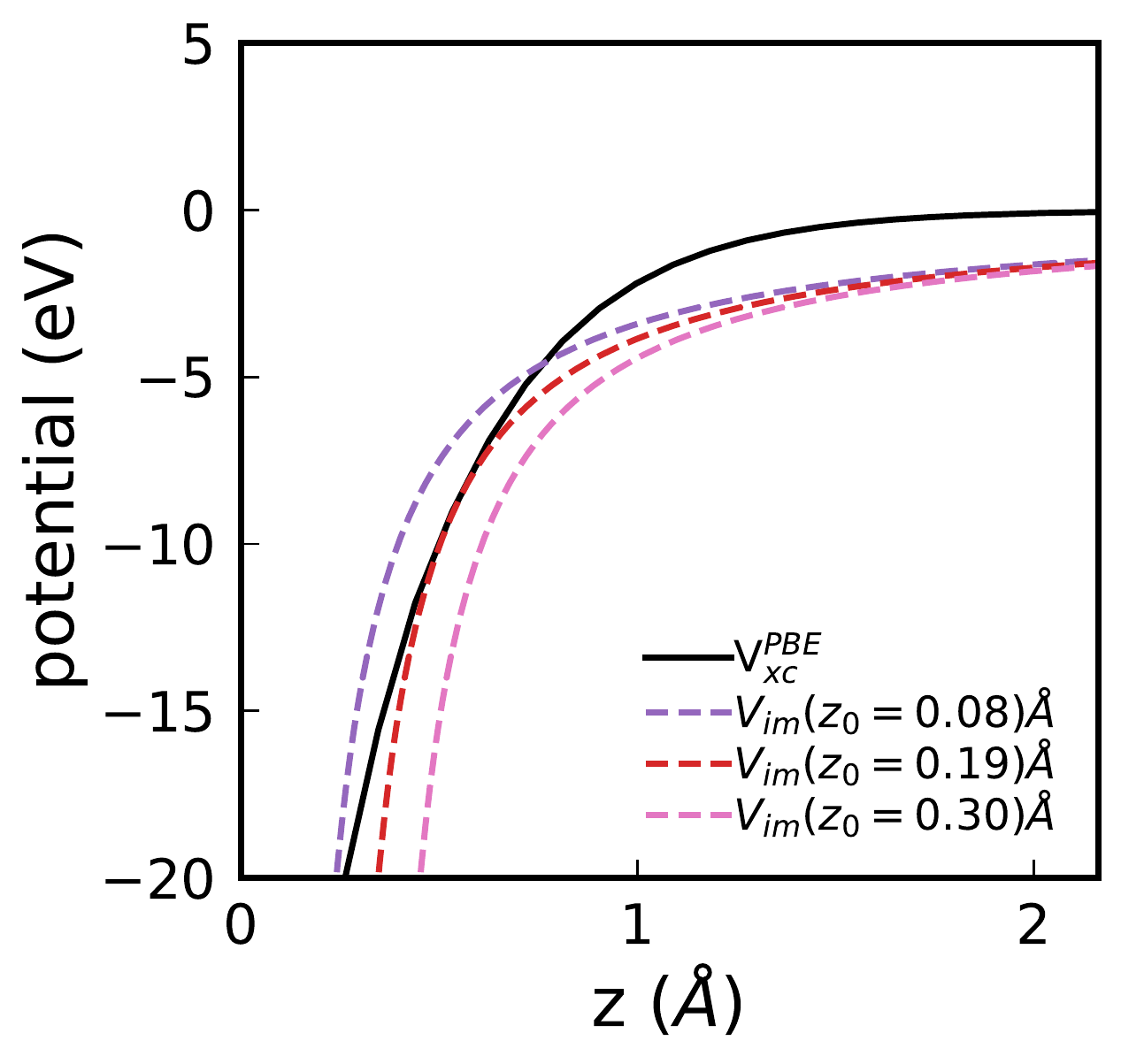}
	\caption{$xy$-plane-averaged exchange-correlation potential ($V_{vx}^{PBE}$, black) along $z$ direction obtained from PBE for 2D MoS$_2$. The electrostatic image potential $V_{im}=\frac{1}{4(z-z_0)}\frac{\varepsilon-1}{\varepsilon+1}$ is plotted for different $z_0$ values of 0.08 \AA (purple), 0.19 \AA (red), and 0.30 \AA (pink). $\varepsilon$=14 uses the dielectric constant of bulk MoS$_2$ based on GW calculations~\cite{Lam2012MoS2scGW,Berkelbach2013MoS2}. The final image-plane position is determined when the $V_{vx}^{PBE}$ curve and $V_{im}$ curve has the same tangent, therefore $z_0=0.19$ \AA.}
	\label{figS:ImP}
\end{figure}

The exchange-correlation potential from PBE, denoted as $V_{im}^{PBE}$, although incorrectly decays exponentially far away from the surface, is correct inside the material, and connects seemlessly to the correct, asymptotically decayed image potential outside the surface. Different from metal surfaces, we need to include the dielectric screening effect, therefore the image potential is $V_{im}=\frac{1}{4(z-z_0)}\frac{\varepsilon-1}{\varepsilon+1}$. By computing $V_{im}$ with various values of $z_0$ and plotting with $V_{im}^{PBE}$, as shown in Fig.~\ref{figS:ImP}, image plane position for MoS$_2$ is $z_0=0.19$ \AA, at which the $V_{im}$ and $V_{im}^{PBE}$ curve have the same tangent. 

\section{Energy Levels with Different Substrates}

  \begin{figure}[H]
	\centering
	\includegraphics[width=0.95\textwidth]{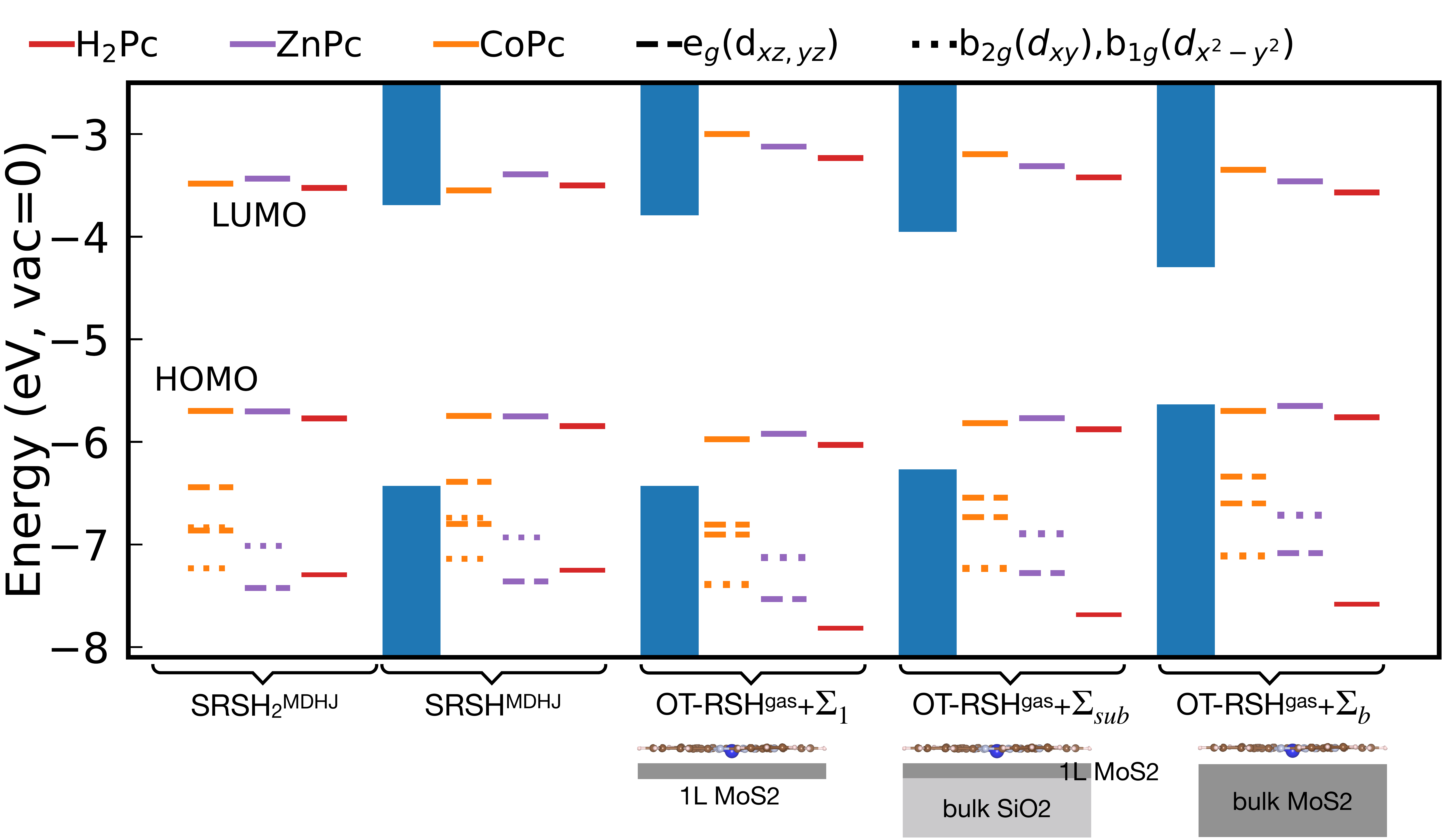}
	\caption{Valence and Conduction bands for MoS$_2$ (blue). Orbital energies for CoPc (orange), ZnPc (purple), and H$_2$Pc (red) at Pc/MoS$_2$ MDHJ from full SRSH calculations for the heterostructure using SRSH$_2^{MDHJ}$ ($\alpha=0.1, \beta=0.234, \gamma=0.140$ Bohr$^{-1}$) and SRSH$^{MDHJ}$ ($\alpha=0.1, \beta=0.5, \gamma=0.05$ Bohr$^{-1}$), 
		OT-RSH$^{gas}+\Sigma_1$ ($\Sigma_{sub}$ and $\Sigma_b$) for energy corrections with dielectric screening using Eq. 1 in the case of molecules on 1L MoS$_2$ (1L/SiO$_2$, and bulk MoS$_2$, respectively). $\Sigma_1$ is the same as in Fig. 3. HOMO/LUMO for Pcs are in solid lines and orbital energies below HOMO are shown as dashed lines for e$_g(d_{xz},d_{yz})$ 
		orbitals and dotted lines for b$_{2g} (d_{xy})$ and b$_{1g} (d_{x^2-y^2})$ orbitals. For MoS$_2$, $\Sigma$ accounts for the dielectric confinement~\cite{Kumagai1989dielectric,Cho2018PRB}, see details in Section~\ref{Sec:2Dsigma}.  }
	\label{figS:EdgeSub}
\end{figure}

 \begin{figure}[H] 
	\centering
	\includegraphics[width=0.95\textwidth]{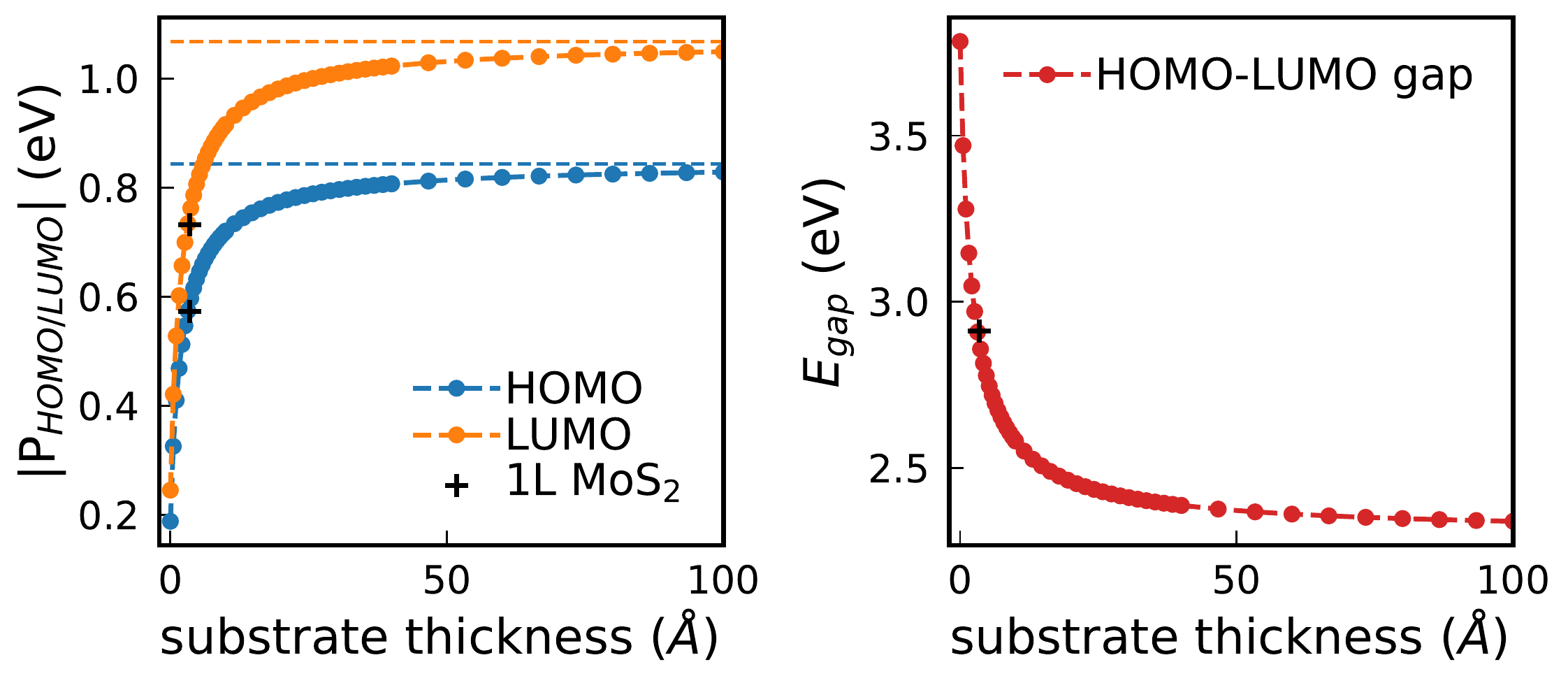}
	\caption{The dielectric screening effect on HOMO and LUMO energies (a) and HOMO-LUMO energy gap (b) of phthalocyanine molecules as a function of MoS$_2$ thickness with $\varepsilon_1=14$.}
	\label{fig:dPthickness}
\end{figure}

\section{DOS and pDOS for MDHJs from PBE}
\addcontentsline{toc}{section}{DOS and pDOS for MDHJs from PBE}
 \begin{figure}[H] 
	\centering
	\includegraphics[width=0.95\textwidth]{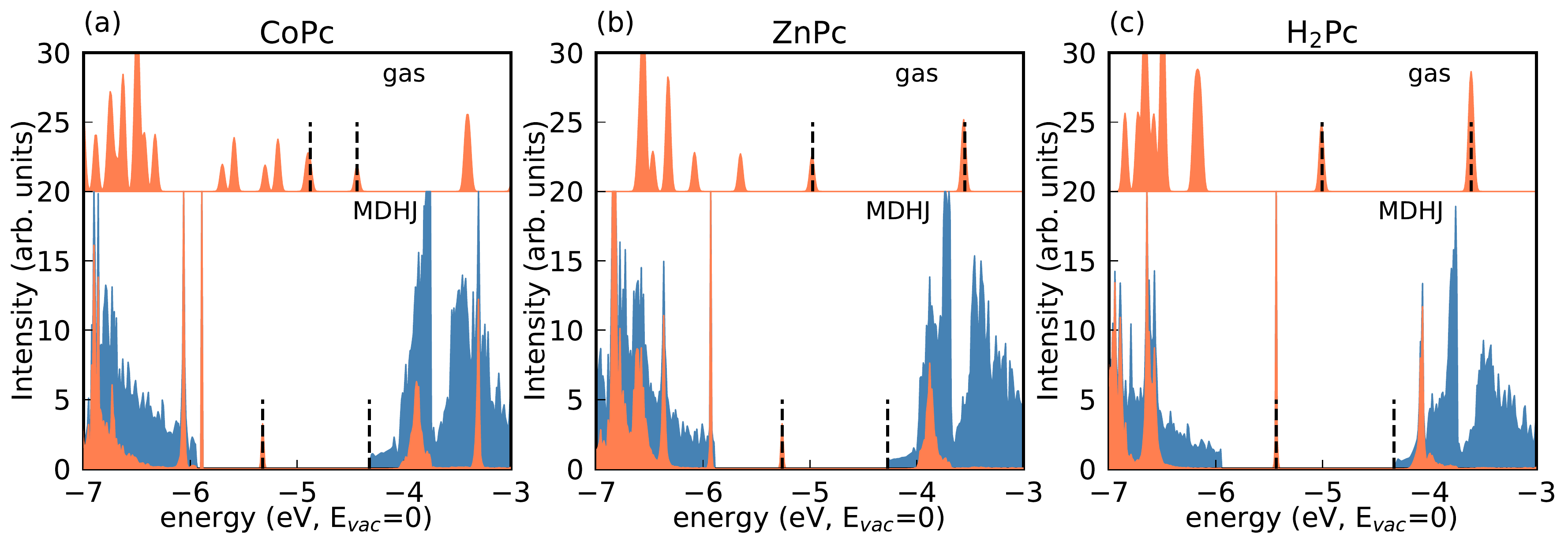}
	\caption{DOS of gas-phase, pDOS of MDHJs for CoPc (a), ZnPc (b) and H$_2$Pc (c), respectively. Dashed lines show positions of VBM, CBM. Results are all from PBE.}
	\label{fig:dospbe}
\end{figure}

 \begin{figure}[H] 
	\centering
	\includegraphics[width=0.6\textwidth]{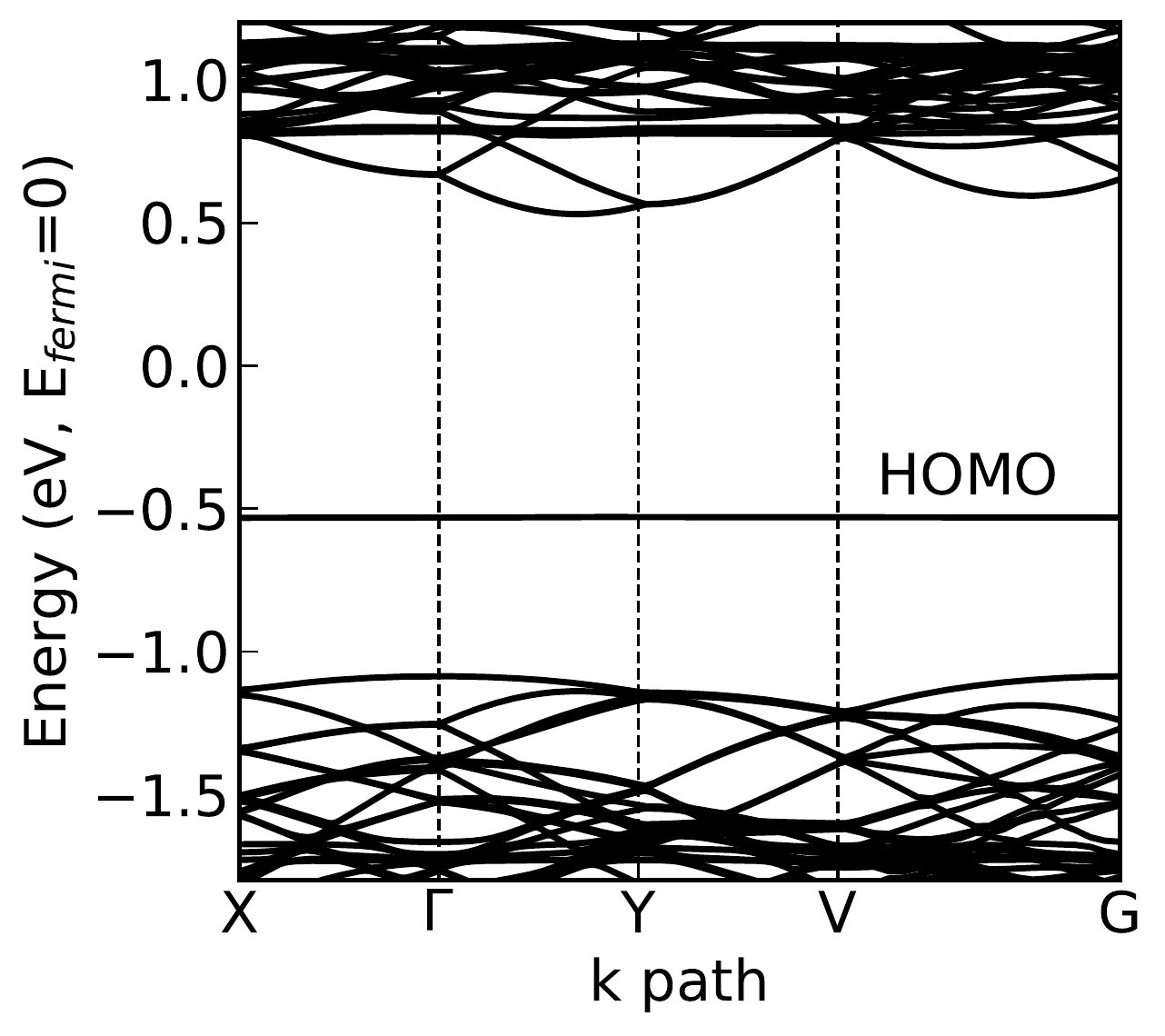}
	\caption{Band structure of H$_2$Pc/MoS$_2$. The horizontal line at about -0.5 eV shows the HOMO energy of H$_2$Pc, same as for gas-phase molecules with no inter-molecular interactions. }
	\label{fig:bnd}
\end{figure}

\section{Electrostatic solution to dielectric screening effects for 2D MoS$_2$} \label{Sec:2Dsigma}
\addcontentsline{toc}{section}{Electrostatic solution to dielectric screening effects}

For 2D monolayer MoS$_2$, here we use a electrostatic model~\cite{Cho2018PRB} to add self-energy corrections to standard DFT-PBE calculations, as well as accounting for dielectric environments.
From bulk to ML MoS$_2$, there are two changes: (1) geometric quantum confinement of carriers and (2) dielectric contrast. For normal DFT calculations from PBE functional, the prior is accounted for but not the later. Therefore we can predict the bandgap of ML MoS$_2$ $E_{g,2D}$ as follows~\cite{Cho2018PRB}:
\begin{align}
E_{g,2D} &= E_{g,bulk}+\Delta E_{g,QF}+\Delta E_{g,\varepsilon} \\
& =  E_{g,bulk}^{exp}+(E_{g,2D}^{PBE}-E_{g,bulk}^{PBE})+\Delta E_{g,\varepsilon}
\label{eq:sigma}
\end{align}
where $E_{g,bulk}^{exp}$, $E_{g,bulk}^{PBE}$, and $E_{g,2D}^{PBE}$ are bandgaps of bulk MoS$_2$ from experiments (1.2 eV)~\cite{Heinz2010PRLMoS2} and PBE calculations, and that of 2D from PBE calculations, respectively. $\Delta E_{g,QF}, \Delta E_{g,\varepsilon}$ are changes of bandgaps due to geometric and dielectric quantum confinement, respectively. Here we apply a self-interaction energy correction $\Sigma$ to the valence- and conduction-band using a electrostatic approximation~\cite{Kumagai1989dielectric,Cho2018PRB} to predict $\Delta E_{g,\varepsilon}$, by considering 2D MoS$_2$ as a homogeneous dielectric slab of $\varepsilon_1$, surrounded by dielectric environments of $\varepsilon_2$ on top and $\varepsilon_3$ below. For free-standing ML MoS$_2$, the environments are vacuum both below and above, therefore $\varepsilon_2=\varepsilon_3=\varepsilon_0=1$ where $\varepsilon_0$ is the dielectric constant of vacuum . For ML MoS$_2$ with surface Pc molecules, $\varepsilon_2=\varepsilon_{Pc}, \varepsilon_3=\varepsilon_0$ where $\varepsilon_{Pc}$ is the dielectric constant of Pc molecular layers, and we use  $\varepsilon_{Pc}$=1.9~\cite{Shi2007PcDielectric}, which is the out-of-plane component of the dielectric tensor for both MPc and H$_2$Pc. $\varepsilon_1$=14~\cite{Lam2012MoS2scGW,Berkelbach2013MoS2} is the dielectric constant for bulk MoS$_2$. The thickness of the dielectric layer of MoS$_2$ is $d=d(MoS_2)+2z_0)$ where $d(MoS_2)=3.13$ \AA\ and $z_0$ are geometric thickness of ML MoS$_2$ and the image plane position normal to the surface, respectively.

\section{Optimized Atomic Coordinates}
\addcontentsline{toc}{section}{Optimized Atomic Coordinates}
Relaxed molecular structures for H2Pc, ZnPc, and CoPc are attached in XYZ format; monolayer MoS$_2$ and  H2Pc/MoS$_2$, ZnPc/MoS$_2$, and CoPc/MoS$_2$ MDHJ structures are attached in crystallographic information file (cif) format.

\bibliography{MPcbib}